\documentclass[12pt,onecolumn]{IEEEtran}

\usepackage{mathrsfs} 
\usepackage{amssymb}
\usepackage{graphicx}
\usepackage{amsmath}
\usepackage[colorlinks,linkcolor=blue,citecolor=red,linktocpage=true]{hyperref}

\usepackage[pagewise]{lineno}

\newcommand{\Tr}{\textrm{\rm Tr}}
\newcommand{\N}{\textrm{\rm N}}
\newcommand{\ord}{\textrm{\rm ord}}

\newcommand{\Occ}{\mathscr{N}}
\newcommand{\CLI}{\mathscr{C}}

\newcommand{\residuo}[2]{(#1 \mbox{ mod } #2)}

\usepackage{cite}

\begin {document}

\def\bbbr{{\rm I\!R}} 

\def\bbbf{{\rm I\!F}}

\def\bbbz{{\mathchoice {\hbox{$\sf\textstyle Z\kern-0.4em Z$}}
{\hbox{$\sf\textstyle Z\kern-0.4em Z$}}
{\hbox{$\sf\scriptstyle Z\kern-0.3em Z$}}
{\hbox{$\sf\scriptscriptstyle Z\kern-0.2em Z$}}}}

\newtheorem{definition}{Definition}
\newtheorem{proposition}{Proposition}
\newtheorem{theorem}{Theorem}

\newtheorem{remark}{Remark}
\newtheorem{example}{Example}

\newenvironment{proof}{\begin{trivlist}\item[]{\em Proof: }}%
{\samepage \hfill{\hbox{\rlap{$\sqcap$}$\sqcup$}}\end{trivlist}}

\title{The complete weight distribution of a family of irreducible cyclic codes of dimension two}

\author{
Gerardo Vega and F\'elix Hern\'andez\thanks{G. Vega and F. Hern\'andez are with the Direcci\'on General de C\'omputo y de Tecnolog\'{\i}as de Informaci\'on y Comunicaci\'on, Uni\-ver\-si\-dad Nacional Aut\'onoma de 
M\'exico, 04510 Ciudad de M\'exico, MEXICO (e-mail: gerardov@unam.mx and felixhdz@ciencias.unam.mx).}\thanks{Manuscript partially supported by PAPIIT-UNAM IN107423.}}
\maketitle


\begin{abstract} 
An important family of codes for data storage systems, cryptography, consumer electronics, and network coding for error control in digital communications are the so-called cyclic codes. This kind of linear codes are also important due to their efficient encoding and decoding algorithms. Because of this, cyclic codes have been studied for many years, however their complete weight distributions are known only for a few cases. The complete weight distribution has a wide range of applications in many research fields as the information it contains is of vital use in practical applications. Unfortunately, obtaining these distributions is in general a very hard problem that normally involves the evaluation of sophisticated exponential sums, which leaves this problem open for most of the cyclic codes. In this paper we determine, for any finite field $\bbbf_q$, the explicit factorization of any polynomial of the form $x^{q+1}-c$, where $c \in \bbbf_{q}^*$. Then we use this result to obtain, without the need to evaluate any kind of exponential sum, the complete weight distributions of a family of irreducible cyclic codes of dimension two over any finite field. As an application of our findings, we employ the complete weight distributions of some irreducible cyclic codes presented here to construct systematic authentication codes, showing that they are optimal or almost optimal. 
\end{abstract}


\section{Introduction}\label{secuno}
The weight distribution of a code is usually investigated on the basis of Hamming weight, under which all the nonzero components of a codeword are regarded as identical. To describe the structure of nonbinary codes in more detail, each nonzero component should be distinguished from the others and this is done by means of the complete weight distribution. In this way, the complete weight distribution of a code, enumerates the codewords by the number of symbols of each kind contained in each codeword. Therefore, the complete weight distribution of a code contains much more information than the Hamming weight distribution. The complete weight distribution has a wide range of applications in many research fields as the information it contains is of vital use in practical applications. For example, the complete weight distributions of linear codes are widely applied in the deception probabilities of certain authentication codes \cite{Ding1, Ding2}. These distributions are also useful in the computation of the Walsh transform of monomial functions over finite fields \cite{Helleseth}, and as pointed out in \cite{Blake}, the complete weight distribution of Reed-Solomon codes could be helpful in soft decision decoding. For this reason, the determination of the complete weight distributions of linear and cyclic codes over finite fields has received a great deal of attention in recent years (see for example \cite{Bae, Chan, Li-Bae, Li-Yue, Yang1, Yang2,Yang3, Zheng}).

Unfortunately, determining the complete weight distribution is an even harder problem than obtaining the Hamming weight distribution, which is often referred to simply as the weight distribution. In fact, obtaining these distributions is in general a very hard problem that normally involves the evaluation of sophisticated exponential sums (see for example \cite{Bae, Heng, Li-Yue}), which leaves this problem open for most of the cyclic codes. In \cite{Shi} and \cite{Vega1}, the weight distribution of any irreducible cyclic code of dimension two is determined, showing that all those irreducible cyclic codes have at most two nonzero weights. This result also shows, contrary to what is asserted in \cite{Rao}, that all the two-weight irreducible cyclic codes of dimension two are semiprimitive (see for example \cite{Schmidt}). A similar result for the complete weight distribution of irreducible cyclic codes of dimension two is until now an open problem. Contributing towards a solution of this problem was what encouraged this study. To this end we determine, for any finite field $\bbbf_q$, the explicit factorization of any polynomial of the form $x^{q+1}-c$, where $c \in \bbbf_{q}^*$. Then we use this result to obtain, without the need to evaluate any kind of exponential sum, the complete weight distributions of a family of irreducible cyclic codes of dimension two over any finite field. As an application of our findings, we employ the generic construction of Cartesian authentication codes presented in \cite{Ding1} and the explicit complete weight distributions of some irreducible cyclic codes presented here to construct systematic authentication codes, showing that they are optimal or almost optimal.

This paper is structured as follows: In Section \ref{secdos}, we establish notation and recall key definitions. To ensure the paper is relatively self-contained, we also recall the result that determines the weight distributions of all irreducible cyclic codes of dimension two. Section \ref{sectres} is devoted to presenting some preliminary results. Particularly, we identify all irreducible polynomials of degree two that have the same constant term. In Section \ref{seccuatro}, we determine the explicit factorization of any polynomial of the form $x^{q+1}-c$, where $c \in \bbbf_{q}^*$. Then, in Section \ref{seccinco}, we use the irreducible factors found in Section \ref{seccuatro}, to obtain the complete weight distributions of a family of irreducible cyclic codes of dimension two over any finite field. In Section \ref{secseis} we recall the generic construction of Cartesian authentication codes and apply such construction to some irreducible cyclic codes studied here, to obtain authentication codes that are optimal or almost optimal. Finally, Section \ref{conclusiones} is devoted to conclusions.

\section{Notation, definitions, and already-known results}\label{secdos}
First of all, we set for this section and for the rest of this work the following:

\noindent
{\bf Notation.} For integers $v$ and $w$, with $\gcd(v,w)=1$, $\ord_v(w)$ will denote the {\em multiplicative order} of $w$ modulo $v$. In this work, for positive integers $i$, $m$ and $z$ with $m>1$ and $0\leq z<m$, when we write $z=\residuo{i}{m}$, instead of $z \equiv i \!\pmod{m}$, we refer that $z$ is equal to the remainder that results when $i$ is divided by $m$. By using $q$ we denote a power of a prime number. From now on, $\gamma$ will denote a primitive element of $\bbbf_{q^2}$ and we fix $\alpha=\gamma^{q+1}$ as primitive element of $\bbbf_{q}$. Let $N$ be an integer such that $N|(q^2-1)$. For $i=0,1,\cdots,N-1$, we define ${\cal C}_i^{(N,q^2)}:=\gamma^i \langle \gamma^N \rangle$, where $\langle \gamma^N \rangle$ denotes the subgroup of $\bbbf_{q^2}^*$ generated by $\gamma^N$. The cosets ${\cal C}_i^{(N,q^2)}$ are called the {\em cyclotomic classes} of order $N$ in $\bbbf_{q^2}$. Note that if $i\geq N$, then ${\cal C}_i^{(N,q^2)}:=\gamma^i \langle \gamma^N \rangle=\gamma^{\residuo{i}{N}} \langle \gamma^N \rangle={\cal C}_{\residuo{i}{N}} ^{(N,q^2)}$. Therefore, as a pragmatic approach the subscript $i$, in the cyclotomic class ${\cal C}_i^{(N,q^2)}$, will be seen modulo $N$. By using ``$\Tr_{\bbbf_{q^2}/\bbbf_q}$'' and ``$\N_{\bbbf_{q^2}/\bbbf_q}$'' we will denote, respectively, the {\em trace mapping} and the {\em norm mapping} from $\bbbf_{q^2}$ to $\bbbf_q$. Let $t \in \bbbf_{q}$, and let $V=(v_0,v_1,\cdots,v_{m-1})$ be a vector of length $m$ over $\bbbf_{q}$. We define {\em the number of occurrences of the symbol $t$ in $V$}, denoted by $\Occ(t,V)$, as the number of times that $t$ occurs as an entry in the vector $V$. That is:

$$\Occ(t,V\!=\!(v_0,v_1,\cdots,v_{m-1}))\!:=\!|\{ i \:|\: t=v_i, \:0\leq i < m\}|\;.$$

For any vector $V=(v_0,v_1,\cdots,v_{m-1})$, the circular shift to the right of $V$, denoted by $\sigma(V)$, is defined as 

$$\sigma(V):=(v_{m-1},v_0,v_1,\cdots,v_{m-2})\;,$$

\noindent
and for any integer, $i\geq 0$, the recursively application, for $i$ times, of $\sigma$ to the vector $V$ will be denoted by $\sigma^i(V)$. For example, $\sigma(3,2,2)=(2,3,2)$, $\sigma^0(3,2,2)=(3,2,2)$ and $\sigma^3(2,3,0,2,4)=(0,2,4,2,3)$.

Let $n\geq 1$ be an integer. We recall that the {\em weight enumerator} or {\em Hamming weight enumerator} of a linear code $\CLI$ of length $n$ is defined as the polynomial $1+\sum_{j=1}^{n} A_j z^j$, where $A_j\: (1 \leq j \leq n)$ denotes the number of codewords with Hamming weight $j$ in the code $\CLI$. The sequence $1,\{A_1, A_2,\cdots,A_n\}$ is called the {\em weight distribution} or {\em Hamming weight distribution} of the code. An $M$-\textit{weight} code is a code such that the cardinality of the set of nonzero weights is $M$, that is $M=|\{i:A_i\neq0,i=1,2,\ldots,n\}|$. 

\subsection{The complete weight distribution and the irreducible cyclic codes of dimension 2}
In a similar way, we give now the definition of the complete weight distribution of a linear code:

\begin{definition}\label{defcero} 
Let $\CLI$ be a linear code of length $n$ over $\bbbf_{q}$. Denote the $q$ elements of $\bbbf_{q}$ by using $u_{-1}=0$, $u_0=\alpha^0=1,u_1=\alpha^1,\cdots,u_{q-2}=\alpha^{q-2}$. By denoting $\mathbb{N}_0:=\mathbb{N} \cup \{0\}$, we define the {\em complete weight} of a vector $\mathbf{v}=(v_0,v_1,\cdots,v_{n-1}) \in \bbbf_{q}^n$, as the vector $w_{\rm cplt}(\mathbf{v}):=(f_0,f_1,\cdots,f_{q-2}) \in \mathbb{N}_0^{q-1}$, where $f_i$ ($0 \leq i \leq q-2$) is the number of components $v_j$ ($0 \leq j < n$) of $\mathbf{v}$ that are equal to $u_i=\alpha^i$, that is $f_i=\Occ(\alpha^i,\mathbf{v})$. In addition, for a vector ${\vec f}=(f_0,f_1,\cdots,f_{q-2}) \in \mathbb{N}_0^{q-1}$ we denote by $Z^{\vec f}$ the monomial in the $q-1$ variables $(z_0,z_1,\cdots,z_{q-2})$ given by 

$$Z^{\vec f}:=z_0^{f_0}z_1^{f_1}\cdots z_{q-2}^{f_{q-2}}\;.$$

\noindent
Now, for a linear code $\CLI$ of length $n$ over $\bbbf_{q}$, we define the {\em set of complete nonzero weights} of ${\CLI}$, $W_{\CLI}$, by means of: 

$$W_{\CLI}:=\{w_{\rm cplt}(\mathbf{c}) \:|\: \mathbf{c} \mbox{ is a nonzero codeword in } {\CLI}\}\;,$$

\noindent
and for each complete nonzero weight ${\vec w} \in W_{\CLI}$, we define its frequency, $A_{\vec w}$, as:

$$A_{\vec w}:=|\{\:\mathbf{c} \in {\CLI} \:|\: w_{\rm cplt}(\mathbf{c})={\vec w}\:\}|\;.$$

\noindent
The sequence $1,\{A_{\vec w}\}_{{\vec w} \in W_{\CLI}}$ is called the {\em complete weight distribution} of the linear code $\CLI$, whereas the polynomial  

\begin{equation}\label{eqCWE}
\mbox{\rm CWE}_{\CLI}(Z):=1+\sum_{{\vec w} \in W_{\CLI}} A_{\vec w} \:Z^{\vec w} \;,
\end{equation}

\noindent
is called its {\em complete weight enumerator}.
\end{definition}

\begin{remark}\label{remuno}
Let $n$ and $\mathbf{v}$ be as before, and let $f_{-1}:\mathbb{N}_0^{q-1} \to \mathbb{N}_0$ be the function given by

$$f_{-1}(f_0,f_1,\cdots,f_{q-2}):=n-\sum_{i=0}^{q-2}f_i\;.$$

\noindent
Thus observe that $\Occ(0,\mathbf{v})=f_{-1}(w_{\rm cplt}(\mathbf{v}))$. Therefore, it is important to keep in mind that an alternative definition for the complete weight enumerator (see for example \cite[p. 141]{MacWilliams}) is:

$$\mbox{\rm CWE}_{\CLI}(Z):=z_{-1}^n+\sum_{{\vec w} \in W_{\CLI}} A_{\vec w} \:z_{-1}^{f_{-1}({\vec w})}Z^{\vec w} \;.$$

\noindent
For linear codes these two definitions are equivalent and, for the convenience of this work, we are going to use (\ref{eqCWE}). In addition, observe also that (\ref{eqCWE}) coincides with the Hamming weight enumerator when $q=2$ and contains much more information if $q > 2$.
\end{remark}

The following gives an explicit description of an irreducible cyclic code of length $n$ and dimension $\ord_n(q)\leq 2$ over $\bbbf_q$.

\begin{definition}\label{defuno} 
Let $n$ and $N$ be integers such that $nN=q^2-1$. Then the set 

$$\CLI_{N}:= \{\mathbf{c}_{N}(a) \:|\: a \in \bbbf_{q^2} \}\;,$$

\noindent
where

$$\mathbf{c}_{N}(a):=(\Tr_{\bbbf_{q^2}/\bbbf_q}(a\gamma^{N k}))_{k=0}^{n-1}\;,$$

\noindent
is an {\em irreducible cyclic code} of length $n$ and dimension $\ord_n(q)\leq 2$ over $\bbbf_q$. 
\end{definition} 

\subsection{The weight distribution of all irreducible cyclic codes of dimension 2}
We end this section by recalling the following result, which characterizes the weight distributions of all irreducible cyclic codes of dimension 2 in terms of their lengths. This characterization shows that all irreducible cyclic codes of dimension 2 have at most two nonzero weights. 

\begin{theorem}\label{teoseis}
\cite[Theorem 7]{Vega1} Let $n$ and $N$ be as before. Fix $u=\gcd(q+1,N)$. Let $\CLI_{N}$ be the irreducible cyclic code of length $n$ over $\bbbf_{q}$. Then:

\begin{enumerate}
\item[\rm (A)] If $u=1$, then $\CLI_{N}$ is an $[n,2]$ one-weight irreducible cyclic code, whose nonzero weight is $\frac{n}{q+1} q$.

\item[\rm (B)] If $2 \leq u < q+1$, then $\CLI_{N}$ is an $[n,2,\frac{n(q+1-u)}{q+1}]$ two-weight irreducible cyclic code whose weight enumerator is 

\[ 1+\frac{(q^2-1)}{u}z^{\frac{n(q+1-u)}{q+1}}+\frac{(q^2-1)(u-1)}{u}z^{n} \; . \] 

\item[\rm (C)] If $u=q+1$, then $\CLI_{N}$ is an $[n,1]$ one-weight irreducible cyclic code, whose nonzero weight is $n$ (which is equivalent to a repetition code of length $n$).

\end{enumerate}
\end{theorem}

\section{Preliminary results}\label{sectres}
The following result, which is likely already known, identifies all irreducible polynomials of degree two that have the same constant term. For completeness, we include its proof here.

\begin{proposition}\label{prouno}
Let $c \in \bbbf_q^*$ and let $f(x)\neq 1$ be a monic irreducible polynomial in $\bbbf_q[x]$. Let $0\leq i < q-1$ be the unique integer such that $c=\alpha^i$. Then 

\begin{equation}\label{eqcero}
x^{q+1}-c=\prod_{r \in {\cal C}_i^{(q-1,q^2)}}(x-r)\;,
\end{equation}

\noindent
and if $f(x)$ is a divisor of $x^{q+1}-c$ then $f(x)$ is either of the form $f(x)=x^2 \pm bx+c$ or $f(x)=x \pm e$, for some elements $b,e \in \bbbf_q$. More specifically, $f(x)=x^2 \pm bx+c$ iff there exists $r \in {\cal C}_i^{(q-1,q^2)} \setminus \bbbf_q^*$ such that  $b= \Tr_{\bbbf_{q^2}/\bbbf_q}(r)$ and $c=\N_{\bbbf_{q^2}/\bbbf_q}(r)$. Alternatively, $f(x)=x \pm e$ iff $e \in {\cal C}_i^{(q-1,q^2)} \cap \bbbf_q^*$, $c$ is a square in $\bbbf_q$ and $e^2=c$. 
\end{proposition}

\begin{proof}
Let $r \in {\cal C}_i^{(q-1,q^2)}$ and suppose that $r=\gamma^{i+j(q-1)}$ for some $j$. Note that $\N_{\bbbf_{q^2}/\bbbf_q}(r)=r^{q+1}=\gamma^{(q+1)i+(q^2-1)j}=\gamma^{(q+1)i}=\alpha^i=c$. Consequently, (\ref{eqcero}) holds, because $|{\cal C}_i^{(q-1,q^2)}|=q+1$ and $r^{q+1}-c=0$ iff $r \in {\cal C}_i^{(q-1,q^2)}$. Hence we have $(x^{q+1}-c)|(x^{q^2-1}-1)$ and if $f(x) | (x^{q+1}-c)$, then $1\leq \deg(f(x))\leq 2$. Now, if $q$ is even then $-r=r$ and if $q$ is odd then $-r=\gamma^{\frac{q+1}{2}(q-1)}\gamma^{i+j(q-1)}=\gamma^{i+(j+\frac{q+1}{2})(q-1)}\in {\cal C}_i^{(q-1,q^2)}$. On the other hand, whether or not $q$ is even, we have 

$$r^q=\gamma^{i+i(q-1)+j(q-1)(q+1-1)}=\gamma^{i+i(q-1)-j(q-1)}=\gamma^{i+(i-j)(q-1)} \in {\cal C}_i^{(q-1,q^2)}\;.$$ 

\noindent
Therefore $r\in {\cal C}_i^{(q-1,q^2)}$ iff $-r,r^q \in {\cal C}_i^{(q-1,q^2)}$. Clearly, $(x+r)(x+r^q)=x^2+\Tr_{\bbbf_{q^2}/\bbbf_q}(r)x+\N_{\bbbf_{q^2}/\bbbf_q}(r)$ and $(x-r)(x-r^q)=x^2-\Tr_{\bbbf_{q^2}/\bbbf_q}(r)x+\N_{\bbbf_{q^2}/\bbbf_q}(r)$. Hence, the polynomial $f(x)=x^2 \pm bx+c$ is an irreducible divisor of $x^{q+1}-c$ iff there exists $r \in {\cal C}_i^{(q-1,q^2)} \setminus \bbbf_q^*$ such that $b= \Tr_{\bbbf_{q^2}/\bbbf_q}(r)$ and $c=\N_{\bbbf_{q^2}/\bbbf_q}(r)$. Finally, if $e \in {\cal C}_i^{(q-1,q^2)} \cap \bbbf_q^*$ then $0=e^{q+1}-c=e^2-c$ and $f(x)=x \pm e$ is an irreducible divisor of $x^{q+1}-c$. 
\end{proof}

\begin{remark}\label{remtres}
Let $N=q-1$. By Definition \ref{defuno}, the length of the irreducible cyclic code $\CLI_{q-1}$ is $n=\frac{q^2-1}{N}=q+1$ and if $a \in {\cal C}_i^{(q-1,q^2)}$, then

\begin{equation}\label{eqcuatro}
\mathbf{c}_{q-1}(a)=(\Tr_{\bbbf_{q^2}/\bbbf_q}(a\gamma^{(q-1)k}))_{k=0}^{q} \in \CLI_{q-1}\;,
\end{equation}

\noindent
and observe also that if $q$ is even then $|{\cal C}_i^{(q-1,q^2)} \cap \bbbf_q^*|=1$. In fact if $q$ is even and $i=2j$, for $0\leq j \leq \frac{q}{2}-1$, then $\alpha^{j}=\gamma^{2j}\gamma^{(q-1)j}\in {\cal C}_{2j}^{(q-1,q^2)}$ and if $i=2j+1$, for $0\leq j \leq \frac{q}{2}-2$, then $\alpha^{\frac{q}{2}+j}=\gamma^{2j+1}\gamma^{(q-1)(\frac{q}{2}+j+1)}\in {\cal C}_{2j+1}^{(q-1,q^2)}$. Alternatively, if $q$ is odd then $|{\cal C}_i^{(q-1,q^2)} \cap \bbbf_q^*|=2$ if $i$ is even and $|{\cal C}_i^{(q-1,q^2)} \cap \bbbf_q^*|=0$ if $i$ is odd. In fact if $q$ is odd and $i=2j$, for $0\leq j \leq \frac{q-1}{2}$, then $\alpha^{j}=\gamma^{2j}\gamma^{(q-1)j}\in {\cal C}_{2j}^{(q-1,q^2)}$ and $-\alpha^{j}=\gamma^{2j}\gamma^{(q-1)(\frac{q+1}{2}+j)}\in {\cal C}_{2j}^{(q-1,q^2)}$. Now, since $a \in {\cal C}_i^{(q-1,q^2)}$, it is important to note that ${\cal C}_i^{(q-1,q^2)}=\{a\gamma^{(q-1)k}\:|\:0\leq k<q+1\}$. Let $h \in \bbbf_q$. Thus, due to Proposition \ref{prouno} and (\ref{eqcuatro}), we have 

$$\Occ(h,\mathbf{c}_{q-1}(a))=\left\{ \begin{array}{cl}
		\;2 & \mbox{ if there exist } r,r^q \in {\cal C}_i^{(q-1,q^2)} \setminus \bbbf_q^* \\
		    & \mbox{ such that } \Tr_{\bbbf_{q^2}/\bbbf_q}(r)=h , \\
\vspace{-3.5mm}
\\		    
		\;1 & \mbox{ if there exist } e,-e \in {\cal C}_i^{(q-1,q^2)} \cap \bbbf_q^* \\
		    & \mbox{ such that } \Tr_{\bbbf_{q^2}/\bbbf_q}(e)=h , \\
\vspace{-3.5mm}
\\		    
		\;0 & \mbox{ if there does not exist } r \in {\cal C}_i^{(q-1,q^2)} \\
		    & \mbox{ such that } \Tr_{\bbbf_{q^2}/\bbbf_q}(r)=h . \\
			\end{array}
\right .$$

\noindent
Particularly, if $q$ is even then $\Occ(0,\mathbf{c}_{q-1}(a))=1$ and there must exist a unique $e \in {\cal C}_i^{(q-1,q^2)} \cap \bbbf_q^*$ such that $\Tr_{\bbbf_{q^2}/\bbbf_q}(e)=0$. Alternatively, if $q$ is odd and $\Occ(h,\mathbf{c}_{q-1}(a))=1$, then $i$ must be even and there must exist two unique elements $e,-e \in {\cal C}_i^{(q-1,q^2)} \cap \bbbf_q^*$ such that $\Tr_{\bbbf_{q^2}/\bbbf_q}(e)=2e=h$ and $\Tr_{\bbbf_{q^2}/\bbbf_q}(-e)=-2e=-h$. Thus, in any case, the conclusion here is that $0 \leq \Occ(h,\mathbf{c}_{q-1}(a)) \leq 2$ and there exists at least one $u\in \bbbf_q$ such that $\Occ(u,\mathbf{c}_{q-1}(a))=2$.
\end{remark}

\section{Factorization of $x^{q+1}-c$}\label{seccuatro}

Given an element $a\in {\cal C}_i^{(q-1,q^2)}$, the goal is to identify those elements $h\in \bbbf_q$ for which $\Occ(h,\mathbf{c}_{q-1}(a))>0$. By Remark \ref{remtres}, and as detailed below, such elements $h$ are related to the linear coefficient of the irreducible polynomials of degree two over the finite field $\bbbf_q$ and therefore we need a simple criterion to figure out whether a polynomial of degree two is irreducible.

\subsection{An easy way to determine when some particular polynomials of degree two are irreducible}
At this point, we are interested in determining whether a polynomial of the form $x^2 \pm\alpha^i x+\alpha^{v} \in \bbbf_{q}[x]$ is irreducible, with $0\leq i<q-1$ and $v \in \{0,1\}$. To this end, and if $q$ is even, we define the sets:

\begin{eqnarray}
R&:=&\{\:0 \leq i < q-1\:|\:\alpha^{i}=\alpha^{k+1}+\alpha^{q-1-k}, \mbox{ with } 0\leq k<\frac{q}{2}-1\:\}\;,\;\;\;\mbox{ and} \nonumber \\
I&:=&\{\:0,1,2,\cdots,q-2\:\} \setminus R\;. \nonumber
\end{eqnarray}

\noindent
Thus, if $0\leq i<q-1$, then $x^2+\alpha^i x+\alpha=(x+\alpha^{k+1})(x+\alpha^{q-1-k})$, for some $0\leq k<\frac{q}{2}-1$, iff $i\in R$, and therefore, $x^2+\alpha^i x+\alpha$ is irreducible iff $i\in I$. Now, if $q$ is odd, we first need to observe that $x^2+\alpha^i x+\alpha^{v}$ is irreducible iff $x^2-\alpha^i x+\alpha^{v}$ is irreducible, and since $-1=\alpha^{\frac{q-1}{2}}$, we can therefore assume, WLOG, that $0\leq i<\frac{q-1}{2}$. In this context, we recall that $\residuo{i}{\frac{q-1}{2}}$ refers to the remainder that results when $i$ is divided by $\frac{q-1}{2}$. In consequence, in the case when $q$ is odd, we define the sets:

\begin{eqnarray}
R_0&:=&\{\:\residuo{i}{\frac{q-1}{2}}\:|\:\alpha^{i}=\alpha^{k}+\alpha^{q-1-k}\;,\mbox{ with } 0\leq k<\left\lceil \frac{q-1}{4} \right\rceil\:\}\;,\nonumber \\
I_0&:=&\{\:0,1,2,\cdots,\frac{q-1}{2}-1\:\} \setminus R_0\;,\nonumber \\
R_1&:=&\{\:\residuo{i}{\frac{q-1}{2}}\:|\:\alpha^{i}=\alpha^{k+1}+\alpha^{q-1-k}\;,\mbox{ with } 0\leq k<\left\lfloor \frac{q-1}{4} \right\rfloor\:\}\;,\;\;\;\mbox{ and} \nonumber \\
I_1&:=&\{\:0,1,2,\cdots,\frac{q-1}{2}-1\:\} \setminus R_1\;, \nonumber
\end{eqnarray}

\noindent
where, for any $y\in \bbbr$, $\lceil y \rceil$ denotes the smallest integer greater than or equal to $y$ and $\lfloor y \rfloor$ denotes the largest integer less than or equal to $y$. Thus, if $0\leq i<\frac{q-1}{2}$, then either $x^2+\alpha^i x+1=(x+\alpha^{k})(x+\alpha^{q-1-k})$ or $x^2-\alpha^i x+1=(x+\alpha^{k})(x+\alpha^{q-1-k})$, for some $0\leq k<\lceil \frac{q-1}{4} \rceil$, iff $i \in R_0$, and therefore, for any integer $i$, $x^2+\alpha^i x+1$ is irreducible iff $\residuo{i}{\frac{q-1}{2}}\in I_0$. Similarly, if $0\leq i<\frac{q-1}{2}$, then either $x^2+\alpha^i x+\alpha=(x+\alpha^{k+1})(x+\alpha^{q-1-k})$ or $x^2-\alpha^i x+\alpha=(x+\alpha^{k+1})(x+\alpha^{q-1-k})$, for some $0\leq k<\lfloor \frac{q-1}{4} \rfloor$, iff $i \in R_1$, and therefore, for any integer $i$, $x^2+\alpha^i x+\alpha$ is irreducible iff $\residuo{i}{\frac{q-1}{2}}\in I_1$.

\subsection{The factorization}
Let $v\in \{0,1\}$. In the previous subsection, we determined when a polynomial of the form $x^2 + \alpha^{i}x+\alpha^{v}$ is irreducible, where we fixed $v=1$ if $q$ is even. The aim of this subsection is to generalize this result to any polynomial of degree $2$, and as a by-product, we find the explicit factorization of the polynomial $x^{q+1}-c$, where $c \in \bbbf_{q}^*$. So, in the case when $q$ is odd, we fix $\delta \in \{0,1\}$ in such a way that $\delta=1$ if $q \equiv 3 \pmod{4}$ and $0$ otherwise. Let $v,w\in \{0,1\}$ and, for integers $j$ and $l$, define the following polynomials over $\bbbf_q$:

$$\eta_v^{(2j+v)}(x):=\left\{ \begin{array}{cl}
		\;(x+\alpha^{j})(x-\alpha^{j}) & \mbox{ if } v=0, \\
\\
		\;1 & \mbox{ otherwise,}
			\end{array}
\right .$$

$$\mu_w^{(l)}(x):=\left\{ \begin{array}{cl}
		\;(x^2+\alpha^{l}) & \mbox{ if } w=1, \\
\\
		\;1 & \mbox{ otherwise.}
			\end{array}
\right .$$

With the previous notations and definitions, we can now present an explicit factorization of any polynomial of the form $x^{q+1}-c$, where $c \in \bbbf_{q}^*$: 

\begin{theorem}\label{MiClas1}
If $q$ is even then, for $0\leq j <q-1$, 

\begin{equation}\label{eq1Teo1}
x^{q+1}+\alpha^{2j+1}=(x+(\alpha^{2j+1})^{1/2})\prod_{i \in I}(x^2+\alpha^{i+j} x+\alpha^{2j+1})\;.
\end{equation}

\noindent
On the other hand, if $q$ is odd then, for $0\leq j <\frac{q-1}{2}$ and $v \in \{0,1\}$,

\begin{equation}\label{eq2Teo1}
x^{q+1}-\alpha^{2j+v}=\eta_v^{(2j+v)}(x)\mu_{\delta\oplus v}^{(2j+v)}(x)\prod_{i \in I_v}(x^2+\alpha^{i+j} x+\alpha^{2j+v})(x^2-\alpha^{i+j} x+\alpha^{2j+v})\;,
\end{equation}

\noindent
where ``$\oplus$'' is the usual XOR operation between two bits.
\end{theorem}

\begin{proof}
Firstly, recall that if $q$ is even then any element $c \in \bbbf_q^*$ has a unique square root and therefore the polynomial $x^2+c$ is always reducible. On the contrary, if $q$ is odd then an element $c \in \bbbf_q^*$ is a square iff $c=\alpha^{2j}$, for some $0\leq j < \frac{q-1}{2}$ and the two square roots of $c$ are $\alpha^j$ and $-\alpha^j$. Additionally, note that the polynomial $x^2+\alpha^{2j}$ is irreducible iff $\delta=1$, while the polynomial $x^2+\alpha^{2j+1}$ is irreducible iff $\delta=0$. Now, due to Proposition \ref{prouno} and the way the sets $I$, $I_0$ and $I_1$ were constructed, we have that if $q$ is even then

\begin{equation}\label{equno}
x^{q+1}+\alpha=(x+\alpha^{1/2})\prod_{i \in I}(x^2+\alpha^{i} x+\alpha)\:=\!\!\!\!\!\prod_{r \in {\cal C}_1^{(q-1,q^2)}}(x-r)\;,
\end{equation}

\noindent
and if $q$ is odd, we have instead

\begin{eqnarray}\label{eqdos}
x^{q+1}-1&=&(x+1)(x-1)\mu_{\delta}^{(0)}(x)\prod_{i \in I_0}(x^2+\alpha^{i} x+1)(x^2-\alpha^{i} x+1)\:=\!\!\!\!\!\prod_{r \in {\cal C}_0^{(q-1,q^2)}}(x-r)\;, \\ \label{eqtres}
x^{q+1}-\alpha&=&\mu_{\delta\oplus 1}^{(1)}(x)\prod_{i \in I_1}(x^2+\alpha^{i} x+\alpha)(x^2-\alpha^{i} x+\alpha)\:=\!\!\!\!\!\prod_{r \in {\cal C}_1^{(q-1,q^2)}}(x-r)\;. 
\end{eqnarray}

\noindent
Clearly, if $q$ is even then $(x+(\alpha^{2j+1})^{1/2})|(x^{q+1}+\alpha^{2j+1})$, for any $0\leq j<q-1$. Alternatively, if $q$ is odd then $(x+\alpha^j)(x-\alpha^j)|(x^{q+1}-\alpha^{2j})$, for any $0\leq j<\frac{q-1}{2}$, and the polynomial $x^2+\alpha^{2j+v}$ is an irreducible divisor of $x^{q+1}-\alpha^{2j+v}$ iff $\delta\oplus v=1$. 

Now, we need to prove that $x^2 + \alpha^{i}x+\alpha^{v}$ is irreducible iff $x^2 + \alpha^{i+j}x+\alpha^{2j+v}$ is irreducible. Clearly, owing to the previous equalities and by fixing $v=1$ when $q$ is even, we know that such a polynomial is irreducible if $j=0$. On the contrary, for $j>0$ assume that $x^2 + \alpha^{i+j}x+\alpha^{2j+v}$ is reducible. Let $r_1,r_2 \in \bbbf_q^*$ such that $(x-r_1)(x-r_2)=x^2 + \alpha^{i+j}x+\alpha^{2j+v}$. Hence, $-r_1-r_2=\alpha^{i+j}$, $r_1r_2=\alpha^{2j+v}$ and $(x-r_1/\alpha^{j})(x-r_2/\alpha^{j})=x^2 + \alpha^{i}x+\alpha^v$, which is a contradiction!

Finally, with $b\neq 0$, suppose that $f(x)=x^2+bx+c \in \bbbf_q[x]$ is irreducible. Thus, if $q$ is even then there must exist a unique $0\leq j<q-1$ such that $c=\alpha^{2j+1}$ and if $q$ is odd then there must exist a unique $0\leq j <\frac{q-1}{2}$ and a unique $v \in \{0,1\}$ such that $c=\alpha^{2j+v}$. Let $0\leq i<q-1$ be the unique integer such that $\alpha^i=b/\alpha^j$. We claim that $x^2 + \alpha^{i}x+\alpha^{v}$ is irreducible, where $v=1$ if $q$ is even. Assume otherwise, and let $r_1,r_2 \in \bbbf_q^*$ such that $(x-r_1)(x-r_2)=x^2 + \alpha^{i}x+\alpha^{v}$. Hence, $-r_1-r_2=\alpha^{i}$, $r_1r_2=\alpha^{v}$ and $(x-\alpha^{j}r_1)(x-\alpha^{j}r_2)=x^2 + \alpha^{i+j}x+\alpha^{2j+v}=x^2+bx+c$, a contradiction! In consequence, with $b\neq 0$, any irreducible polynomial $f(x)=x^2+bx+c \in \bbbf_q[x]$ can be written in the form $f(x)=x^2 + \alpha^{i+j}x+\alpha^{2j+v}$, where $i\in I$ if $q$ is even and $\residuo{i}{\frac{q-1}{2}} \in I_{v}$ if $q$ is odd. 
\end{proof}

\begin{example}\label{ejuno}
$ $
\begin{enumerate}
\item[{\rm (A)}] Let $\bbbf_{8}=\bbbf_2(\alpha)$, with $\alpha^3+\alpha+1=0$. Note that $\langle \alpha \rangle=\bbbf_{8}^*$. Then it is not difficult to obtain that $\frac{q}{2}=4$, $R=\{3,0,2\}$ and therefore $I=\{1,4,5,6\}$. Thus, owing to (\ref{eq1Teo1}) in Theorem \ref{MiClas1}, we have

\vspace{-0.5mm}
{\small
\begin{eqnarray}
(x^9+\alpha^1)&\!\!\!\!\!=\!\!\!\!\!&(x+\alpha^4)(x^2+\alpha^1x+\alpha^1)(x^2+\alpha^4x+\alpha^1)(x^2+\alpha^5x+\alpha^1)(x^2+\alpha^6x+\alpha^1)\;, \nonumber \\
(x^9+\alpha^3)&\!\!\!\!\!=\!\!\!\!\!&(x+\alpha^5)(x^2+\alpha^2x+\alpha^3)(x^2+\alpha^5x+\alpha^3)(x^2+\alpha^6x+\alpha^3)(x^2+\alpha^0x+\alpha^3)\;, \nonumber \\
(x^9+\alpha^5)&\!\!\!\!\!=\!\!\!\!\!&(x+\alpha^6)(x^2+\alpha^3x+\alpha^5)(x^2+\alpha^6x+\alpha^5)(x^2+\alpha^0x+\alpha^5)(x^2+\alpha^1x+\alpha^5)\;, \nonumber \\
(x^9+\alpha^0)&\!\!\!\!\!=\!\!\!\!\!&(x+\alpha^0)(x^2+\alpha^4x+\alpha^0)(x^2+\alpha^0x+\alpha^0)(x^2+\alpha^1x+\alpha^0)(x^2+\alpha^2x+\alpha^0)\;, \nonumber \\
(x^9+\alpha^2)&\!\!\!\!\!=\!\!\!\!\!&(x+\alpha^1)(x^2+\alpha^5x+\alpha^2)(x^2+\alpha^1x+\alpha^2)(x^2+\alpha^2x+\alpha^2)(x^2+\alpha^3x+\alpha^2)\;, \nonumber \\
(x^9+\alpha^4)&\!\!\!\!\!=\!\!\!\!\!&(x+\alpha^2)(x^2+\alpha^6x+\alpha^4)(x^2+\alpha^2x+\alpha^4)(x^2+\alpha^3x+\alpha^4)(x^2+\alpha^4x+\alpha^4)\;, \nonumber \\
(x^9+\alpha^6)&\!\!\!\!\!=\!\!\!\!\!&(x+\alpha^3)(x^2+\alpha^0x+\alpha^6)(x^2+\alpha^3x+\alpha^6)(x^2+\alpha^4x+\alpha^6)(x^2+\alpha^5x+\alpha^6)\;. \nonumber
\end{eqnarray}
}

\item[{\rm (B)}] Let $\bbbf_{9}=\bbbf_3(\alpha)$, with $\alpha^2+\alpha+2=0$. Note that $\langle \alpha \rangle=\bbbf_{9}^*$. Then it is not difficult to obtain that $\frac{q-1}{2}=4$, $-1=\alpha^{4}$, $\lceil \frac{q-1}{4} \rceil=\lfloor \frac{q-1}{4} \rfloor=2$, $\delta=0$, $R_0=\{0,2\}$, $R_1=\{3,0\}$ and therefore $I_0=\{1,3\}$ and $I_1=\{1,2\}$. Thus if $v=0$, and owing to (\ref{eq2Teo1}) in Theorem \ref{MiClas1}, we have

\vspace{-0.5mm}
{\small
\begin{eqnarray}
(x^{10}-\alpha^0)&\!\!\!\!\!=\!\!\!\!\!&(x+\alpha^0)(x+\alpha^4)(x^2+\alpha^1x+\alpha^0)(x^2+\alpha^5x+\alpha^0)(x^2+\alpha^3x+\alpha^0)(x^2+\alpha^7x+\alpha^0)\;, \nonumber \\
(x^{10}-\alpha^2)&\!\!\!\!\!=\!\!\!\!\!&(x+\alpha^1)(x+\alpha^5)(x^2+\alpha^2x+\alpha^2)(x^2+\alpha^6x+\alpha^2)(x^2+\alpha^0x+\alpha^2)(x^2+\alpha^4x+\alpha^2)\;, \nonumber \\
(x^{10}-\alpha^4)&\!\!\!\!\!=\!\!\!\!\!&(x+\alpha^2)(x+\alpha^6)(x^2+\alpha^3x+\alpha^4)(x^2+\alpha^7x+\alpha^4)(x^2+\alpha^1x+\alpha^4)(x^2+\alpha^5x+\alpha^4)\;, \nonumber \\
(x^{10}-\alpha^6)&\!\!\!\!\!=\!\!\!\!\!&(x+\alpha^3)(x+\alpha^7)(x^2+\alpha^0x+\alpha^6)(x^2+\alpha^4x+\alpha^6)(x^2+\alpha^2x+\alpha^6)(x^2+\alpha^6x+\alpha^6)\;, \nonumber
\end{eqnarray}
}

\noindent
and if $v=1$, we have

\vspace{-0.5mm}
{\small
\begin{eqnarray}
(x^{10}-\alpha^1)&\!\!\!\!\!=\!\!\!\!\!&(x^2+\alpha^1)(x^2+\alpha^1x+\alpha^1)(x^2+\alpha^5x+\alpha^1)(x^2+\alpha^2x+\alpha^1)(x^2+\alpha^6x+\alpha^1)\;, \nonumber \\
(x^{10}-\alpha^3)&\!\!\!\!\!=\!\!\!\!\!&(x^2+\alpha^3)(x^2+\alpha^2x+\alpha^3)(x^2+\alpha^6x+\alpha^3)(x^2+\alpha^3x+\alpha^3)(x^2+\alpha^7x+\alpha^3)\;, \nonumber \\
(x^{10}-\alpha^5)&\!\!\!\!\!=\!\!\!\!\!&(x^2+\alpha^5)(x^2+\alpha^3x+\alpha^5)(x^2+\alpha^7x+\alpha^5)(x^2+\alpha^0x+\alpha^5)(x^2+\alpha^4x+\alpha^5)\;, \nonumber \\
(x^{10}-\alpha^7)&\!\!\!\!\!=\!\!\!\!\!&(x^2+\alpha^7)(x^2+\alpha^0x+\alpha^7)(x^2+\alpha^4x+\alpha^7)(x^2+\alpha^1x+\alpha^7)(x^2+\alpha^5x+\alpha^7)\;. \nonumber
\end{eqnarray}
}

\item[{\rm (C)}] Let $(q,\alpha)=(11,2)$ and note that $\langle 2 \rangle=\bbbf_{11}^*$. Then it is not difficult to obtain that $\frac{q-1}{2}=5$, $-1=\alpha^{5}$, $\lceil \frac{q-1}{4} \rceil=3$, $\lfloor \frac{q-1}{4} \rfloor=2$, $\delta=1$, $R_0=\{1,3,2\}$, $R_1=\{3,0\}$ and therefore $I_0=\{0,4\}$ and $I_1=\{1,2,4\}$. Thus if $v=0$, and owing to (\ref{eq2Teo1}) in Theorem \ref{MiClas1}, we have

{\small
\begin{eqnarray}
(x^{12}-\alpha^0)&\!\!\!\!\!=\!\!\!\!\!&(x+\alpha^0)(x+\alpha^5)(x^2+\alpha^0)(x^2+\alpha^0x+\alpha^0)(x^2+\alpha^5x+\alpha^0)(x^2+\alpha^4x+\alpha^0)(x^2+\alpha^9x+\alpha^0)\;, \nonumber \\
(x^{12}-\alpha^2)&\!\!\!\!\!=\!\!\!\!\!&(x+\alpha^1)(x+\alpha^6)(x^2+\alpha^2)(x^2+\alpha^1x+\alpha^2)(x^2+\alpha^6x+\alpha^2)(x^2+\alpha^0x+\alpha^2)(x^2+\alpha^5x+\alpha^2)\;, \nonumber \\
(x^{12}-\alpha^4)&\!\!\!\!\!=\!\!\!\!\!&(x+\alpha^2)(x+\alpha^7)(x^2+\alpha^4)(x^2+\alpha^2x+\alpha^4)(x^2+\alpha^7x+\alpha^4)(x^2+\alpha^1x+\alpha^4)(x^2+\alpha^6x+\alpha^4)\;, \nonumber \\
(x^{12}-\alpha^6)&\!\!\!\!\!=\!\!\!\!\!&(x+\alpha^3)(x+\alpha^8)(x^2+\alpha^6)(x^2+\alpha^3x+\alpha^6)(x^2+\alpha^8x+\alpha^6)(x^2+\alpha^2x+\alpha^6)(x^2+\alpha^7x+\alpha^6)\;, \nonumber \\
(x^{12}-\alpha^8)&\!\!\!\!\!=\!\!\!\!\!&(x+\alpha^4)(x+\alpha^9)(x^2+\alpha^8)(x^2+\alpha^4x+\alpha^8)(x^2+\alpha^9x+\alpha^8)(x^2+\alpha^3x+\alpha^8)(x^2+\alpha^8x+\alpha^8)\;, \nonumber
\end{eqnarray}
}

\vspace{-0.8mm}
\noindent
and if $v=1$, we have

\vspace{-0.9mm}
{\small
\begin{eqnarray}
(x^{12}-\alpha^1)&\!\!\!\!\!=\!\!\!\!\!&(x^2+\alpha^1x+\alpha^1)(x^2+\alpha^6x+\alpha^1)(x^2+\alpha^2x+\alpha^1)(x^2+\alpha^7x+\alpha^1)(x^2+\alpha^4x+\alpha^1)(x^2+\alpha^9x+\alpha^1)\;, \nonumber \\
(x^{12}-\alpha^3)&\!\!\!\!\!=\!\!\!\!\!&(x^2+\alpha^2x+\alpha^3)(x^2+\alpha^7x+\alpha^3)(x^2+\alpha^3x+\alpha^3)(x^2+\alpha^8x+\alpha^3)(x^2+\alpha^0x+\alpha^3)(x^2+\alpha^5x+\alpha^3)\;, \nonumber \\
(x^{12}-\alpha^5)&\!\!\!\!\!=\!\!\!\!\!&(x^2+\alpha^3x+\alpha^5)(x^2+\alpha^8x+\alpha^5)(x^2+\alpha^4x+\alpha^5)(x^2+\alpha^9x+\alpha^5)(x^2+\alpha^1x+\alpha^5)(x^2+\alpha^6x+\alpha^5)\;, \nonumber \\
(x^{12}-\alpha^7)&\!\!\!\!\!=\!\!\!\!\!&(x^2+\alpha^4x+\alpha^7)(x^2+\alpha^9x+\alpha^7)(x^2+\alpha^0x+\alpha^7)(x^2+\alpha^5x+\alpha^7)(x^2+\alpha^2x+\alpha^7)(x^2+\alpha^7x+\alpha^7)\;, \nonumber \\
(x^{12}-\alpha^9)&\!\!\!\!\!=\!\!\!\!\!&(x^2+\alpha^0x+\alpha^9)(x^2+\alpha^5x+\alpha^9)(x^2+\alpha^1x+\alpha^9)(x^2+\alpha^6x+\alpha^9)(x^2+\alpha^3x+\alpha^9)(x^2+\alpha^8x+\alpha^9)\;. \nonumber
\end{eqnarray}
}
\end{enumerate}
\end{example}

\section{The complete weight distributions of a family of irreducible cyclic codes of dimension two}\label{seccinco}

In the sequel of the paper, we are interested in determining the complete weight distribution for the family of irreducible cyclic codes, $\CLI_{N}$, for which $N|(q-1)$. As detailed below, the key to achieving this goal is the construction of three vectors. 

\subsection{The three vectors $W$, $W_1$, and $W_2$}
We first fix $N=q-1$ and therefore the length of $\CLI_{q-1}$ is $n=\frac{q^2-1}{q-1}=q+1$. Suppose that $q$ is even and assume that $a \in {\cal C}_1^{(q-1,q^2)}$. Thus, owing to (\ref{equno}) and Remark \ref{remtres}, $\Occ(\alpha^i,\mathbf{c}_{q-1}(a))=2$ if $i\in I$, $\Occ(0,\mathbf{c}_{q-1}(a))=1$, and $\Occ(\alpha^i,\mathbf{c}_{q-1}(a))=0$ if $i \not\in I$, for $0\leq i<q-1$. Therefore, by defining $W=(w_0,w_1,\cdots,w_{q-2})$ as the integer vector of length $q-1$ whose entries are given by

$$w_i=\left\{ \begin{array}{cl}
		\;2\; & \mbox{ if } i \in I \\
		\;0\; & \mbox{ otherwise}
			\end{array}
\right .\;,$$

\noindent
we have that $w_{\rm cplt}(\mathbf{c}_{q-1}(a))=W$. In the case when $q$ odd, we first fix, $0\leq s<\frac{q-1}{2}$, to be the unique integer such that either $\alpha^s=2=\Tr_{\bbbf_{q^2}/\bbbf_q}(1)$ or $\alpha^s=-2=-\Tr_{\bbbf_{q^2}/\bbbf_q}(1)$. Here we want to recall that $-1=\alpha^{\frac{q-1}{2}}$ and therefore $x^2+\alpha^i+1$ is irreducible iff $x^2+\alpha^{\frac{q-1}{2}+i}+1$ is irreducible. Now, assume that $a \in {\cal C}_0^{(q-1,q^2)}$. Thus, owing to (\ref{eqdos}) and Remark \ref{remtres}, $\Occ(\alpha^i,\mathbf{c}_{q-1}(a))=2$ if $\residuo{i}{\frac{q-1}{2}} \in I_0$, $\Occ(\alpha^s,\mathbf{c}_{q-1}(a))=1$, and $\Occ(\alpha^i,\mathbf{c}_{q-1}(a))=0$ if $\residuo{i}{\frac{q-1}{2}} \not\in I_0 \cup \{s\}$. Therefore, by defining $W_0=(w_{0,0},w_{0,1},\cdots,w_{0,\frac{q-1}{2}-1})$ as the integer vector of length $\frac{q-1}{2}$ whose entries are given by

$$w_{0,i}=\left\{ \begin{array}{cl}
		\;2\; & \mbox{ if } i \in I_0 \\
		\;1\; & \mbox{ if } i = s \\
		\;0\; & \mbox{ otherwise}
			\end{array}
\right .\;,$$

\noindent
we have that $w_{\rm cplt}(\mathbf{c}_{q-1}(a))=W_0|W_0$, where the operator ``$|$'' stands for the vector concatenation. Similarly, if $a \in {\cal C}_1^{(q-1,q^2)}$ then, owing to (\ref{eqtres}) and Remark \ref{remtres}, $\Occ(\alpha^i,\mathbf{c}_{q-1}(a))=2$ if $\residuo{i}{\frac{q-1}{2}}\in I_1$ and $\Occ(\alpha^i,\mathbf{c}_{q-1}(a))=0$ if $\residuo{i}{\frac{q-1}{2}} \not\in I_1$. Therefore, by defining $W_1=(w_{1,0},w_{1,1},\cdots,w_{1,\frac{q-1}{2}-1})$ as the integer vector of length $\frac{q-1}{2}$ whose entries are given by

$$w_{1,i}=\left\{ \begin{array}{cl}
		\;2\; & \mbox{ if } i \in I_1 \\
		\;0\; & \mbox{ otherwise}
			\end{array}
\right .\;,$$

\noindent
we have that $w_{\rm cplt}(\mathbf{c}_{q-1}(a))=W_1|W_1$.

\begin{example}\label{ejdos}
$ $
\begin{enumerate}
\item[{\rm (A)}] Let $q=8$. In Example \ref{ejuno} {\rm (A)} we found $I=\{1,4,5,6\}$. Therefore $W=(0,2,0,0,2,2,2)$ and if $a \in {\cal C}_1^{(q-1,q^2)}$ then $w_{\rm cplt}(\mathbf{c}_{q-1}(a))=W$. 

\item[{\rm (B)}] Let $q=9$. In Example \ref{ejuno} {\rm (B)} we found $I_0=\{1,3\}$ and $I_1=\{1,2\}$. Since $2=-1=\alpha^4$, $-2=1=\alpha^0$ and $s=0$. Therefore $W_0=(1,2,0,2)$ and $W_1=(0,2,2,0)$. Furthermore, 

$$w_{\rm cplt}(\mathbf{c}_{q-1}(a))=\left\{ \begin{array}{cl}
		\;W_0|W_0=(1,2,0,2,1,2,0,2)\; & \mbox{ if } a \in {\cal C}_0^{(q-1,q^2)} \\
		\;W_1|W_1=(0,2,2,0,0,2,2,0)\; & \mbox{ if } a \in {\cal C}_1^{(q-1,q^2)}
			\end{array}
\right .\;.$$

\item[{\rm (C)}] Let $q=11$. In Example \ref{ejuno} {\rm (C)} we found $I_0=\{0,4\}$ and $I_1=\{1,2,4\}$. Since $2=\alpha^1$, $s=1$. Therefore $W_0=(2,1,0,0,2)$ and $W_1=(0,2,2,0,2)$. Furthermore,

$$w_{\rm cplt}(\mathbf{c}_{q-1}(a))=\left\{ \begin{array}{cl}
		\;W_0|W_0=(2,1,0,0,2,2,1,0,0,2)\; & \mbox{ if } a \in {\cal C}_0^{(q-1,q^2)} \\
		\;W_1|W_1=(0,2,2,0,2,0,2,2,0,2)\; & \mbox{ if } a \in {\cal C}_1^{(q-1,q^2)}
			\end{array}
\right .\;.$$
\end{enumerate}
\end{example}

\subsection{The complete weight of any codeword in $\CLI_{q-1}$}
Up to this point we are able to determine, by means of vectors $W$, $W_0$ and $W_1$, the complete weight of a codeword $\mathbf{c}_{q-1}(a) \in \CLI_{q-1}$ when $a \in {\cal C}_1^{(q-1,q^2)}$, if $q$ is even, and $a \in {\cal C}_0^{(q-1,q^2)} \cup {\cal C}_1^{(q-1,q^2)}$, if $q$ is odd. Thus, it remains to determine the complete weights for the other possible values of the finite field element $a \in \bbbf_{q^2}^*$. The following result not only gives a solution to this, but even more importantly it shows that such complete weights can also be obtained by means of the same vectors $W$, $W_0$ and $W_1$.

\begin{proposition}\label{prodos}
Let $W$, $W_0$, and $W_1$ be as before. If $q$ is even, then, for any $b \in {\cal C}_{2j+1}^{(q-1,q^2)}$ with $0\leq j<q-1$, we have

\begin{equation}\label{pr2eq1}
w_{\rm cplt}(\mathbf{c}_{q-1}(b))=\sigma^j(W)\;.
\end{equation}

\noindent
On the other hand, if $q$ is odd, then, for any two elements $b \in {\cal C}_{2j}^{(q-1,q^2)}$ and $b' \in {\cal C}_{2j+1}^{(q-1,q^2)}$ with $0\leq j<\frac{q-1}{2}$, we have

\begin{eqnarray}\label{pr2eq2}
w_{\rm cplt}(\mathbf{c}_{q-1}(b))&=&\sigma^j(W_0)|\sigma^j(W_0)\;,  \\
\label{pr2eq3}
w_{\rm cplt}(\mathbf{c}_{q-1}(b'))&=&\sigma^j(W_1)|\sigma^j(W_1)\;. 
\end{eqnarray}
\end{proposition}

\begin{proof}
Suppose that $q$ is even and let $b \in {\cal C}_{2j+1}^{(q-1,q^2)}$ with $0\leq j<q-1$. From our previous discussion, we know $w_{\rm cplt}(\mathbf{c}_{q-1}(a))=W$ for any $a \in {\cal C}_1^{(q-1,q^2)}$. Now, due to the proof of Theorem \ref{MiClas1}, $\residuo{i}{q-1} \in I$ iff $x^2 + \alpha^{i}x+\alpha$ is irreducible iff $x^2 + \alpha^{i+j}x+\alpha^{2j+1}$ is irreducible. Consequently, $\Occ(\alpha^k,\mathbf{c}_{q-1}(b))=2$ iff $\residuo{k}{q-1} \in j+I:=\{\residuo{j+i}{q-1} \:|\: i\in I\}$ and $\Occ(\alpha^k,\mathbf{c}_{q-1}(b))=0$ if $\residuo{k}{q-1} \not\in j+I$. Therefore, by defining $Y=(y_0,y_1,\cdots,y_{q-2})$ as the integer vector of length $q-1$ whose entries are given by

$$y_i=\left\{ \begin{array}{cl}
		\;2\; & \mbox{ if } i \in j+I \\
		\;0\; & \mbox{ otherwise}
			\end{array}
\right .\;,$$

\noindent
we have $w_{\rm cplt}(\mathbf{c}_{q-1}(b))=Y=\sigma^j(W)$. In the case when $q$ is odd, we again fix, $0\leq s<\frac{q-1}{2}$, to be the unique integer such that either $\alpha^s=2=\Tr_{\bbbf_{q^2}/\bbbf_q}(1)$ or $\alpha^s=-2=-\Tr_{\bbbf_{q^2}/\bbbf_q}(1)$. Assume that $b \in {\cal C}_{2j}^{(q-1,q^2)}$, for some $0\leq j<\frac{q-1}{2}$. Since $\alpha^j=\gamma^{2j}\gamma^{(q-1)j} \in {\cal C}_{2j}^{(q-1,q^2)}$, note that either $\Tr_{\bbbf_{q^2}/\bbbf_q}(\alpha^j)=\alpha^{s+j}$ or $\Tr_{\bbbf_{q^2}/\bbbf_q}(\alpha^j)=-\alpha^{s+j}$. Thus, in like manner, $\residuo{i}{\frac{q-1}{2}} \in I_0$ iff $x^2 + \alpha^{i}x+1$ is irreducible iff $x^2 + \alpha^{i+j}x+\alpha^{2j}$ is irreducible. Consequently, $\Occ(\alpha^k,\mathbf{c}_{q-1}(b))=2$ iff $\residuo{k}{\frac{q-1}{2}} \in j+I_0:=\{\residuo{j+i}{\frac{q-1}{2}} \:|\: i\in I_0\}$, $\Occ(\alpha^k,\mathbf{c}_{q-1}(b))=1$ if $\residuo{k}{\frac{q-1}{2}} =\residuo{s+j}{\frac{q-1}{2}}$, and $\Occ(\alpha^k,\mathbf{c}_{q-1}(b))=0$ if $\residuo{k}{\frac{q-1}{2}} \not\in (j+I_0) \cup \{\residuo{s+j}{\frac{q-1}{2}}\}$. Therefore, by defining $Y=(y_0,y_1,\cdots,y_{\frac{q-1}{2}-1})$ as the integer vector of length $\frac{q-1}{2}$ whose entries are given by

$$y_i=\left\{ \begin{array}{cl}
		\;2\; & \mbox{ if } i \in j+I_0 \\
		\;1\; & \mbox{ if } i = \residuo{s+j}{\frac{q-1}{2}} \\
		\;0\; & \mbox{ otherwise}
			\end{array}
\right .\;,$$

\noindent
we have $w_{\rm cplt}(\mathbf{c}_{q-1}(b))=Y|Y=\sigma^j(W_0)|\sigma^j(W_0)$. In a quite similar way, it is easy to see that if $b' \in {\cal C}_{2j+1}^{(q-1,q^2)}$, for some $0\leq j<\frac{q-1}{2}$, then $w_{\rm cplt}(\mathbf{c}_{q-1}(b'))=\sigma^j(W_1)|\sigma^j(W_1)$. 
\end{proof}

\subsection{The functions ${\cal G}$, ${\cal R}$, and ${\cal R}_k$}
As we will see later on, the vectors $W$, $W_0$ and $W_1$ are the key to obtaining the complete weight distribution of any irreducible cyclic code $\CLI_{N}$, with $N|(q-1)$. However, before that, we need to introduce some new notations. Thus, for positive integers $k$ and $l$, such that $k|(q-1)$ and $l|k$, we define the functions ${\cal G}:\bbbz^{k} \times \bbbz \to \bbbz^l$, ${\cal R}:\bbbz^l \to \bbbz^{q-1}$, and ${\cal R}_k:\bbbz^l \to \bbbz^{k}$ given by

\begin{eqnarray}
{\cal G}((v_0,v_1,\cdots,v_{k-1}),l)\!&:=&\sum_{i=0}^{\frac{k}{l}-1}(v_{il},v_{il+1},\cdots,v_{il+l-1})\;, \nonumber \\
{\cal R}((v_0,v_1,\cdots,v_{l-1}))\!&:=&\overbrace{(v_0,v_1,\cdots,v_{l-1})|(v_0,v_1,\cdots,v_{l-1})|\cdots|(v_0,v_1,\cdots,v_{l-1})}^{\frac{q-1}{l} \mbox{ times}}\;, \nonumber \\
{\cal R}_k((v_0,v_1,\cdots,v_{l-1}))\!&:=&\overbrace{(v_0,v_1,\cdots,v_{l-1})|(v_0,v_1,\cdots,v_{l-1})|\cdots|(v_0,v_1,\cdots,v_{l-1})}^{\frac{k}{l} \mbox{ times}}\;, \nonumber
\end{eqnarray}

\noindent
where the previous sum is over the vectors of length $l$ with integer components and, again, the operator ``$|$'' stands for the vector concatenation. 

There exist some interesting relationships among the functions $\sigma$, ${\cal G}$, ${\cal R}$, and ${\cal R}_k$:

\begin{remark}\label{remrel}
Let $k$ and $l$ be as before. Let $V\in \bbbz^{k}$, $Y\in \bbbz^{l}$. Then observe that

$$\sum_{m=0}^{\frac{k}{l}-1}\sigma^{lm}(V)={\cal R}_k({\cal G}(V,l))\;\;\;\mbox{ and}$$
$${\cal R}(Y)=\overbrace{{\cal R}_k(Y)|{\cal R}_k(Y)|\cdots|{\cal R}_k(Y)}^{\frac{q-1}{k} \mbox{ times}}\;.$$
\end{remark}

For example, if $q=13$, $k=6$, $l=2$ and $V=(2,2,0,1,0,2)$, then ${\cal G}(V,6)=V$ whereas ${\cal G}(V,2)=(2,2)+(0,1)+(0,2)=(2,5)$. Thus ${\cal R}({\cal G}(V,6))=(2,2,0,1,0,2,2,2,0,1,0,2)$, $\frac{k}{l}=3$, 

\begin{eqnarray}
\sum_{m=0}^{2}\sigma^{2m}(V)&=&{\cal R}_k({\cal G}(V,2))={\cal G}(V,2)|{\cal G}(V,2)|{\cal G}(V,2)=(2,5,2,5,2,5)\;,\;\;\mbox{ and} \nonumber \\
{\cal R}({\cal G}(V,2))&=&{\cal R}_k({\cal G}(V,2))|{\cal R}_k({\cal G}(V,2))=(2,5,2,5,2,5,2,5,2,5,2,5)\;. \nonumber
\end{eqnarray}

For our vectors $W$, $W_0$, and $W_1$ we have the following:

\begin{proposition}\label{prorel}
Let $W$, $W_0$, and $W_1$ be as before, and let $N$ and $t$ be positive integers such that $N|(q-1)$ and $t=\frac{q-1}{N}$. Let $j$ be an integer such that $0\leq j<\frac{N}{2}$ if $N$ is even and $0\leq j<N$ otherwise. Additionally, let $e$ be a positive integer. Then,

\begin{equation}\label{rm3eq1}
\sum_{m=0}^{t-1} \sigma^{j+Nm}(W)={\cal R}(\sigma^j({\cal G}(W,N)))\;,
\end{equation}

\noindent
while if $q$ is odd and $N$ is even, then, for $v \in \{0,1\}$, we have:

\begin{equation}\label{rm3eq2}
\sum_{m=0}^{t-1} \sigma^{j+\frac{N}{2}m}(W_v|W_v)={\cal R}(\sigma^j({\cal G}(W_v,\frac{N}{2})))\;, 
\end{equation}

\noindent
and if $q$ is odd and $N$ is odd, then we have:

\begin{equation}\label{rm3eq3}
\sum_{m=0}^{\frac{t}{2}-1} \sigma^{j+Nm}(W_0|W_0)+\sigma^{j+e+Nm}(W_1|W_1)={\cal R}(\sigma^j({\cal G}(W_0+\sigma^e(W_1),N)))\;.
\end{equation}
\end{proposition}

\begin{proof} Suppose that $q$ is odd and $N$ is odd. Then

\begin{eqnarray}
\sum_{m=0}^{\frac{t}{2}-1} \sigma^{j+Nm}(W_0|W_0)+\sigma^{j+e+Nm}(W_1|W_1)\!\!\!&=&\!\!\!\sum_{m=0}^{\frac{t}{2}-1}\sigma^{Nm}(\sigma^{j}(W_0+\sigma^e(W_1))|\sigma^{j}(W_0+\sigma^e(W_1)))\;, \nonumber \\
\!\!\!&=&\!\!\!\sum_{m=0}^{\frac{t}{2}-1}\sigma^{Nm}(\sigma^{j}(W_0+\sigma^e(W_1)))|\sum_{m=0}^{\frac{t}{2}-1}\sigma^{Nm}(\sigma^{j}(W_0+\sigma^e(W_1)))\;. \nonumber
\end{eqnarray} 

\noindent
Since $N$ is odd, $N|\frac{q-1}{2}$. Let $k=\frac{q-1}{2}$ and note that $\frac{k}{N}=\frac{t}{2}$. Then, owing to Remark \ref{remrel},

$$\sum_{m=0}^{\frac{t}{2}-1}\sigma^{Nm}(\sigma^{j}(W_0+\sigma^e(W_1)))={\cal R}_k({\cal G}(\sigma^{j}(W_0+\sigma^e(W_1)),N))\;.$$

\noindent
Therefore

\begin{eqnarray}
\sum_{m=0}^{\frac{t}{2}-1} \sigma^{j+Nm}(W_0|W_0)+\sigma^{j+e+Nm}(W_1|W_1)\!\!\!&=&\!\!\!{\cal R}_k({\cal G}(\sigma^{j}(W_0+\sigma^e(W_1)),N))|{\cal R}_k({\cal G}(\sigma^{j}(W_0+\sigma^e(W_1)),N))\;, \nonumber \\
\!\!\!&=&\!\!\!{\cal R}({\cal G}(\sigma^{j}(W_0+\sigma^e(W_1)),N))\;, \nonumber \\
\!\!\!&=&\!\!\!{\cal R}(\sigma^j({\cal G}(W_0+\sigma^e(W_1),N)))\;. \nonumber
\end{eqnarray} 

\noindent
In a similar way, (\ref{rm3eq1}) and (\ref{rm3eq2}) can be proved. 
\end{proof}

\begin{example}\label{ejtres}
$ $
\begin{enumerate}
\item[{\rm (A)}] Let $\bbbf_{16}=\bbbf_2(\alpha)$, with $\alpha^4+\alpha+1=0$. Note that $\langle \alpha \rangle=\bbbf_{16}^*$. Then it is not difficult to obtain that $\frac{q}{2}=8$, $R=\{4,13,8,6,3,7,0\}$, $I=\{1,2,5,9,10,11,12,14\}$ and therefore 

\begin{equation}\label{eq16}
W=(0,2,2,0,0,2,0,0,0,2,2,2,2,0,2)\;.
\end{equation}

\noindent
Let $(N,j)=(5,2)$, thus $t=3$ and

\begin{eqnarray}
\sigma^{2+5(0)}(W)&=&\sigma^{\;2}(W)=(0,2,0,2,2,0,0,2,0,0,0,2,2,2,2)\;, \nonumber \\
\sigma^{2+5(1)}(W)&=&\sigma^{\;7}(W)=(0,2,2,2,2,0,2,0,2,2,0,0,2,0,0)\;, \nonumber \\
\sigma^{2+5(2)}(W)&=&\sigma^{ 12}(W)\!=(0,0,2,0,0,0,2,2,2,2,0,2,0,2,2)\;. \nonumber
\end{eqnarray}

\noindent
On the other hand, note that ${\cal G}(W,N)=(0,2,2,0,0)+(2,0,0,0,2)+(2,2,2,0,2)=(4,4,4,0,4)$, $\sigma^2({\cal G}(W,N))=(0,4,4,4,4)$  and clearly

$$\sigma^{2}(W)+\sigma^{7}(W)+\sigma^{12}(W)=(0,4,4,4,4,0,4,4,4,4,0,4,4,4,4)={\cal R}(0,4,4,4,4)\;,$$

\noindent
which is in accordance with (\ref{rm3eq1}).

\item[{\rm (B)}] Let $(q,N)=(9,4)$. Thus $t=2$, $\frac{N}{2}=2$, and from Example \ref{ejdos} {\rm (B)}, $W_0=(1,2,0,2)$. With $j=1$, we have

\begin{eqnarray}
\sigma^{1+2(0)}(W_0|W_0)&=&\sigma^{1}(W_0|W_0)=(2,1,2,0,2,1,2,0)\;, \nonumber \\
\sigma^{1+2(1)}(W_0|W_0)&=&\sigma^{3}(W_0|W_0)=(2,0,2,1,2,0,2,1)\;. \nonumber
\end{eqnarray}

\noindent
On the other hand, note that ${\cal G}(W_0,2)=(1,2)+(0,2)=(1,4)$, $\sigma^1(1,4)=(4,1)$ and clearly

$$\sigma^{1}(W_0|W_0)+\sigma^{3}(W_0|W_0)=(4,1,4,1,4,1,4,1)={\cal R}(4,1)\;,$$

\noindent
which is in accordance with (\ref{rm3eq2}).

\item[{\rm (C)}] Let $(q,N,e)=(11,5,2)$. Thus $t/2=1$ and from Example \ref{ejdos} {\rm (C)}, $W_0=(2,1,0,0,2)$ and $W_1=(0,2,2,0,2)$. With $j=3$, we have

\begin{eqnarray}
\sigma^{3+5(0)}(W_0|W_0)&=&\sigma^{3}(W_0|W_0)  =(0,0,2,2,1,0,0,2,2,1)\;, \nonumber \\
\sigma^{3+2+5(0)}(W_1|W_1)&=&\sigma^{5}(W_1|W_1)=(0,2,2,0,2,0,2,2,0,2)\;. \nonumber
\end{eqnarray}

\noindent
On the other hand, ${\cal G}(W_0+\sigma^2(W_1),5)={\cal G}((2,3,0,2,4),5)=(2,3,0,2,4)$, $\sigma^3(2,3,0,2,4)=(0,2,4,2,3)$ and clearly

$$\sigma^{3}(W_0|W_0)+\sigma^{5}(W_1|W_1)=(0,2,4,2,3,0,2,4,2,3)={\cal R}(0,2,4,2,3)\;,$$

\noindent
which is in accordance with (\ref{rm3eq3}).
\end{enumerate}
\end{example}

\subsection{The complete weight enumerator for the irreducible cyclic code $\CLI_{q-1}$}
In terms of the three vectors $W$, $W_0$, and $W_1$, the following result determines the complete weight enumerator for the irreducible cyclic code $\CLI_{q-1}$:

\begin{theorem}\label{teodos}
Let $W$, $W_0$, and $W_1$ be as before. If $q$ is even, then $\CLI_{q-1}$ is a $[q+1,2]$ one-weight irreducible cyclic code whose nonzero weight is $q$ and whose complete weight enumerator is:

$$\mbox{\rm CWE}_{\CLI_{q-1}}(Z)=1+(q+1)\sum_{j=0}^{q-2} \:Z^{\sigma^j(W)} \;,$$

\noindent
and if $q$ is odd, then $\CLI_{q-1}$ is a $[q+1,2]$ two-weight irreducible cyclic code whose nonzero weights are $q-1$ and $q+1$, and whose complete weight enumerator is: 

$$\mbox{\rm CWE}_{\CLI_{q-1}}(Z)=1+(q+1)\sum_{j=0}^{\frac{q-1}{2}-1} \:[\:Z^{{\cal R}(\sigma^j(W_0))} + Z^{{\cal R}(\sigma^j(W_1))}\:]\;.$$
\end{theorem}

\begin{proof}
Clearly $\ord_{q+1}(q)=2$, therefore in accordance with Definition \ref{defuno}, $\CLI_{q-1}$, is a cyclic code of length $n=q+1$ and dimension $2$. Let $u=\gcd(q+1,\frac{q^2-1}{q+1})=\gcd(q+1,q-1)$. Clearly $u=1$ if $q$ is even and $u=2$ if $q$ is odd. Therefore, owing to Theorem \ref{teoseis}, if $q$ is even, then $\CLI_{q-1}$ is a $[q+1,2]$ one-weight irreducible cyclic code whose nonzero weight is $q$ and if $q$ is odd, then $\CLI_{q-1}$ is a $[q+1,2]$ two-weight irreducible cyclic code whose nonzero weights are $q-1$ and $q+1$.   

Let $a\in \bbbf_{q^2}^*$. Thus, by Proposition \ref{prodos}, we have that if $q$ is even then $w_{\rm cplt}(\mathbf{c}_{q-1}(a))=\sigma^j(W)$, for some $0\leq j<q-1$. On the other hand, if $q$ is odd and $a \in {\cal C}_{2j}^{(q-1,q^2)}$ then $w_{\rm cplt}(\mathbf{c}_{q-1}(a))=\sigma^j(W_0)|\sigma^j(W_0)={\cal R}(\sigma^j(W_0))$ and if $a \in {\cal C}_{2j+1}^{(q-1,q^2)}$ then $w_{\rm cplt}(\mathbf{c}_{q-1}(a))=\sigma^j(W_1)|\sigma^j(W_1)={\cal R}(\sigma^j(W_1))$, for some $0\leq j<\frac{q-1}{2}$. Finally, note that if $a \in {\cal C}_i^{(q-1,q^2)}$ then $w_{\rm cplt}(\mathbf{c}_{q-1}(r))=w_{\rm cplt}(\mathbf{c}_{q-1}(a))$, for any $r \in {\cal C}_i^{(q-1,q^2)}$, and therefore the remaining part of the proof follows from the fact that $|{\cal C}_{i}^{(q-1,q^2)}|=q+1$.
\end{proof}

\begin{example}\label{ejcuatro}
$ $
\begin{enumerate}
\item[{\rm (A)}] Let $q=8$. In Example \ref{ejdos} {\rm (A)} we found $W=(0,2,0,0,2,2,2)$. Thus, by Theorem \ref{teodos}, we have that $\CLI_{7}$ is a $[9,2]$ one-weight irreducible cyclic code whose nonzero weight is $8$ and whose complete weight enumerator is

\begin{eqnarray}
\mbox{\rm CWE}_{\CLI_{7}}(Z)=1+9(&&\!\!\!\!\!\!\!\!\!\!z_0^0z_1^2z_2^0z_3^0z_4^2z_5^2z_6^2+z_0^2z_1^0z_2^2z_3^0z_4^0z_5^2z_6^2+z_0^2z_1^2z_2^0z_3^2z_4^0z_5^0z_6^2+z_0^2z_1^2z_2^2z_3^0z_4^2z_5^0z_6^0+\nonumber \\
                                 &&\!\!\!\!\!\!\!\!\!\!z_0^0z_1^2z_2^2z_3^2z_4^0z_5^2z_6^0+z_0^0z_1^0z_2^2z_3^2z_4^2z_5^0z_6^2+z_0^2z_1^0z_2^0z_3^2z_4^2z_5^2z_6^0\:)\;. \nonumber
\end{eqnarray}

\item[{\rm (B)}] Let $q=9$. In Example \ref{ejdos} {\rm (B)} we found $W_0=(1,2,0,2)$ and $W_1=(0,2,2,0)$. Note that $\sigma^1(W_0)=(2,1,2,0)$, $\sigma^2(W_0)=(0,2,1,2)$ and $\sigma^3(W_0)=(2,0,2,1)$ while $\sigma^1(W_1)=(0,0,2,2)$, $\sigma^2(W_1)=(2,0,0,2)$ and $\sigma^3(W_1)=(2,2,0,0)$. Thus, by Theorem \ref{teodos}, we have that $\CLI_{8}$ is a $[10,2]$ two-weight  irreducible cyclic code whose nonzero weights are $8$ and $10$, and whose complete weight enumerator is

\begin{eqnarray}
\mbox{\rm CWE}_{\CLI_{8}}(Z)=1+10(&&\!\!\!\!\!\!\!\!\!\!z_0^1z_1^2z_2^0z_3^2z_4^1z_5^2z_6^0z_7^2+z_0^0z_1^2z_2^2z_3^0z_4^0z_5^2z_6^2z_7^0+\nonumber \\
                                  &&\!\!\!\!\!\!\!\!\!\!z_0^2z_1^1z_2^2z_3^0z_4^2z_5^1z_6^2z_7^0+z_0^0z_1^0z_2^2z_3^2z_4^0z_5^0z_6^2z_7^2+\nonumber \\
                                  &&\!\!\!\!\!\!\!\!\!\!z_0^0z_1^2z_2^1z_3^2z_4^0z_5^2z_6^1z_7^2+z_0^2z_1^0z_2^0z_3^2z_4^2z_5^0z_6^0z_7^2+\nonumber \\
                                  &&\!\!\!\!\!\!\!\!\!\!z_0^2z_1^0z_2^2z_3^1z_4^2z_5^0z_6^2z_7^1+z_0^2z_1^2z_2^0z_3^0z_4^2z_5^2z_6^0z_7^0\:)\;. \nonumber
\end{eqnarray}

\item[{\rm (C)}] Let $q=11$. In Example \ref{ejdos} {\rm (C)} we found $W_0=(2,1,0,0,2)$ and $W_1=(0,2,2,0,2)$. Note that $\sigma^1(W_0)=(2,2,1,0,0)$, $\sigma^2(W_0)=(0,2,2,1,0)$, $\sigma^3(W_0)=(0,0,2,2,1)$ and $\sigma^4(W_0)=(1,0,0,2,2)$ while $\sigma^1(W_1)=(2,0,2,2,0)$, $\sigma^2(W_1)=(0,2,0,2,2)$, $\sigma^3(W_1)=(2,0,2,0,2)$ and $\sigma^4(W_1)=(2,2,0,2,0)$. Thus, by Theorem \ref{teodos}, we have that $\CLI_{10}$ is a $[12,2]$ two-weight irreducible cyclic code whose nonzero weights are $10$ and $12$, and whose complete weight enumerator is

\begin{eqnarray}
\mbox{\rm CWE}_{\CLI_{10}}(Z)=1+12(&&\!\!\!\!\!\!\!\!\!\!z_0^2z_1^1z_2^0z_3^0z_4^2z_5^2z_6^1z_7^0z_8^0z_9^2+z_0^0z_1^2z_2^2z_3^0z_4^2z_5^0z_6^2z_7^2z_8^0z_9^2+\nonumber \\
                                   &&\!\!\!\!\!\!\!\!\!\!z_0^2z_1^2z_2^1z_3^0z_4^0z_5^2z_6^2z_7^1z_8^0z_9^0+z_0^2z_1^0z_2^2z_3^2z_4^0z_5^2z_6^0z_7^2z_8^2z_9^0+\nonumber \\
                                   &&\!\!\!\!\!\!\!\!\!\!z_0^0z_1^2z_2^2z_3^1z_4^0z_5^0z_6^2z_7^2z_8^1z_9^0+z_0^0z_1^2z_2^0z_3^2z_4^2z_5^0z_6^2z_7^0z_8^2z_9^2+\nonumber \\
                                   &&\!\!\!\!\!\!\!\!\!\!z_0^0z_1^0z_2^2z_3^2z_4^1z_5^0z_6^0z_7^2z_8^2z_9^1+z_0^2z_1^0z_2^2z_3^0z_4^2z_5^2z_6^0z_7^2z_8^0z_9^2+\nonumber \\
                                   &&\!\!\!\!\!\!\!\!\!\!z_0^1z_1^0z_2^0z_3^2z_4^2z_5^1z_6^0z_7^0z_8^2z_9^2+z_0^2z_1^2z_2^0z_3^2z_4^0z_5^2z_6^2z_7^0z_8^2z_9^0\:)\;. \nonumber
\end{eqnarray}
\end{enumerate}
\end{example}

\subsection{The complete weight enumerator of the family of irreducible cyclic codes $\CLI_{N}$ with $N|(q-1)$}
We are now interested in extending our previous result to the family of irreducible cyclic codes of length $n$, such that $Nn=q^2-1$ and $N|(q-1)$. So, once again in terms of the three vectors $W$, $W_0$, and $W_1$, and for any $N|(q-1)$, the following result determines the complete weight enumerator for any irreducible cyclic code, $\CLI_N$, of length $n=\frac{q^2-1}{N}$ and dimension $2$ over any finite field $\bbbf_q$.

\begin{theorem}\label{teotres}
Let $W$, $W_0$, and $W_1$ be as before and let $N$ be a positive integer such that $N|(q-1)$. If $N$ is odd, fix $e=\frac{N-1}{2}$. Then $\CLI_{N}$ is a cyclic code of length $n=\frac{q^2-1}{N}$ and dimension $2$. In addition, if $q$ is even, then $\CLI_{N}$ is an $[n,2]$ one-weight irreducible cyclic code whose nonzero weight is $q\frac{q-1}{N}$ and whose complete weight enumerator is:

\begin{equation}\label{teo3eq1}
\mbox{\rm CWE}_{\CLI_N}(Z)=1+n\sum_{j=0}^{N-1} \:Z^{{\cal R}(\sigma^j({\cal G}(W,N)))} \;.
\end{equation}

\noindent
If $q$ is odd and $N$ is even, then $\CLI_{N}$ is an $[n,2]$ two-weight irreducible cyclic code whose nonzero weights are $n-\frac{2n}{q+1}$ and $n$, and whose complete weight enumerator is:

\begin{equation}\label{teo3eq2}
\mbox{\rm CWE}_{\CLI_N}(Z)=1+n\sum_{j=0}^{\frac{N}{2}-1} \:[Z^{{\cal R}(\sigma^j({\cal G}(W_0,\frac{N}{2})))}+Z^{{\cal R}(\sigma^j({\cal G}(W_1,\frac{N}{2})))}]\;,
\end{equation}

\noindent
and if $q$ is odd and $N$ is odd, then $\CLI_{N}$ is an $[n,2]$ one-weight irreducible cyclic code whose nonzero weight is $q\frac{q-1}{N}$ and whose complete weight enumerator is:

\begin{equation}\label{teo3eq3}
\mbox{\rm CWE}_{\CLI_N}(Z)=1+n\sum_{j=0}^{N-1} \:Z^{{\cal R}(\sigma^j({\cal G}(W_0+\sigma^e(W_1),N)))}\;.
\end{equation}
\end{theorem}

\begin{proof}
Since $q+1 \leq n=\frac{q^2-1}{N}$, $\ord_{n}(q)=2$, therefore in accordance with Definition \ref{defuno}, $\CLI_{N}$, is a cyclic code of length $n$ and dimension $2$. Clearly $u=\gcd(q+1,N)=1$ if ether $q$ is even or $q$ is odd and $N$ is odd. On the other hand, $u=2$ if $q$ is odd and $N$ even. Therefore, owing to Theorem \ref{teoseis}, if ether $q$ is even or $q$ is odd and $N$ is odd, then $\CLI_{N}$ is an $[n,2]$ one-weight irreducible cyclic code whose nonzero weight is $\frac{n}{q+1}q$, and if $q$ is odd and $N$ even, then $\CLI_{N}$ is an $[n,2]$ two-weight irreducible cyclic code whose nonzero weights are $n-\frac{2n}{q+1}$ and $n$.   

Fix $t=\frac{q-1}{N}$ and note that, for any integer $l$, we have

$${\cal C}_{l}^{(N,q^2)}=\bigcup_{m=0}^{t-1} {\cal C}_{l+Nm}^{(q-1,q^2)}\;.$$

Assume that $q$ is even and let $a \in {\cal C}_{i}^{(N,q^2)}$, for some $0\leq i<N$. Since $N$ is odd, there must exist a unique integer $0\leq j<N$ such that $i=\residuo{2j+1}{N}$, where we recall here that $\residuo{2j+1}{N}$ is the remainder of the division of $2j+1$ by $N$. Similarly, observe that 

$$\{\: Nm \:|\:0\leq m<t\: \}=\{\: \residuo{2Nm}{q-1} \:|\:0\leq m<t\: \}\;,$$ 

\noindent
thus

$$\{\: \residuo{2j+1+Nm}{q-1} \:|\:0\leq m<t\: \}=\{\: \residuo{2(j+Nm)+1}{q-1} \:|\:0\leq m<t\: \}\;.$$

\noindent
Therefore 

$${\cal C}_{i}^{(N,q^2)}={\cal C}_{2j+1}^{(N,q^2)}=\bigcup_{m=0}^{t-1} {\cal C}_{2j+1+Nm}^{(q-1,q^2)}=\bigcup_{m=0}^{t-1} {\cal C}_{2(j+Nm)+1}^{(q-1,q^2)}\;.$$

\noindent
Let $a_0,a_1,\cdots,a_{t-1} \in {\cal C}_{i}^{(N,q^2)}$ such that $a_m \in {\cal C}_{2(j+Nm)+1}^{(q-1,q^2)}$, for $m=0,1,\cdots,t-1$. But 

\begin{eqnarray}
{\cal C}_{i}^{(N,q^2)}&=&\{\: a\gamma^{N k} \:|\:0\leq k<n=(q+1)t \:\}\;\;\;\mbox{ and } \nonumber \\
{\cal C}_{2(j+Nm)+1}^{(q-1,q^2)}&=&\{\: a_m\gamma^{(q-1) k} \:|\:0\leq k<q+1 \:\}\;,  \nonumber
\end{eqnarray}

\noindent
thus

\begin{eqnarray}
w_{\rm cplt}(\mathbf{c}_{N}(a))&=&\sum_{m=0}^{t-1}w_{\rm cplt}(\mathbf{c}_{q-1}(a_m))\;,  \nonumber \\
&=&\sum_{m=0}^{t-1} \sigma^{j+Nm}(W)\;,  \nonumber \\
&=&{\cal R}(\sigma^j({\cal G}(W,N)))\;, \nonumber
\end{eqnarray}

\noindent
where the second and the third equalities come from (\ref{pr2eq1}) and (\ref{rm3eq1}), respectively. In consequence (\ref{teo3eq1}) now follows from the fact that $0\leq j<N$ and $|{\cal C}_{i}^{(N,q^2)}|=\frac{q^2-1}{N}=n$.

Assume that $q$ is odd and $N$ is even. Let $a \in {\cal C}_{i}^{(N,q^2)}$, for some $0\leq i<N$. Then, there must exist unique integers $0\leq j<\frac{N}{2}$ and $v\in\{0,1\}$ such that $i=2j+v$. Therefore, 

$${\cal C}_{i}^{(N,q^2)}={\cal C}_{2j+v}^{(N,q^2)}=\bigcup_{m=0}^{t-1} {\cal C}_{2j+v+Nm}^{(q-1,q^2)}=\bigcup_{m=0}^{t-1} {\cal C}_{2(j+\frac{N}{2}m)+v}^{(q-1,q^2)}\;.$$

\noindent
Let $a_0,a_1,\cdots,a_{t-1} \in {\cal C}_{i}^{(N,q^2)}$ such that $a_m \in {\cal C}_{2(j+\frac{N}{2}m)+v}^{(q-1,q^2)}$, for $m=0,1,\cdots,t-1$. Thus, in a quite similar way as before, and by using now (\ref{pr2eq2}), (\ref{pr2eq3}), and (\ref{rm3eq2}), we have

$$w_{\rm cplt}(\mathbf{c}_{N}(a))=\sum_{m=0}^{t-1}w_{\rm cplt}(\mathbf{c}_{q-1}(a_m))=\sum_{m=0}^{t-1} \sigma^{j+\frac{N}{2}m}(W_v|W_v)={\cal R}(\sigma^j({\cal G}(W_v,\frac{N}{2})))\;.$$

\noindent
In consequence (\ref{teo3eq2}) now follows from the fact that $0\leq j<\frac{N}{2}$, $v\in\{0,1\}$, and $|{\cal C}_{i}^{(N,q^2)}|=\frac{q^2-1}{N}=n$.

Assume that $q$ and $N$ are odd. Observe that $t$ is even. Let $a \in {\cal C}_{i}^{(N,q^2)}$, for some $0\leq i<N$. Then, there must exist a unique integer $0\leq j<N$ given by 

$$j=\left\{ \begin{array}{cl}
		\;\frac{i}{2}\; & \mbox{ if $i$ is even,} \\
		\\
		\;\frac{i+N}{2}\; & \mbox{ if $i$ is odd.}
			\end{array}
\right .$$

\noindent
Thus, it is not difficult to see that

\begin{eqnarray}
\{\: i+Nm\:|\:0\leq m<t\: \}&=&\{\: 2(j+Nm)\:|\:0\leq m<\frac{t}{2}\: \}\cup \nonumber \\
&&\{\: \residuo{2(j+e+Nm)+1}{q-1}\:|\:0\leq m<\frac{t}{2}\: \}\;,  \nonumber
\end{eqnarray}

\noindent
therefore 

$${\cal C}_{i}^{(N,q^2)}=\bigcup_{m=0}^{t-1} {\cal C}_{i+Nm}^{(q-1,q^2)}=\bigcup_{m=0}^{\frac{t}{2}-1} \left({\cal C}_{2(j+Nm)}^{(q-1,q^2)}\cup {\cal C}_{2(j+e+Nm)+1}^{(q-1,q^2)}\right)\;.$$

\noindent
Let $a_0,a_1,\cdots,a_{\frac{t}{2}-1},a'_0,a'_1,\cdots,a'_{\frac{t}{2}-1} \in {\cal C}_{i}^{(N,q^2)}$ such that $a_m \in {\cal C}_{2(j+Nm)}^{(q-1,q^2)}$ and $a'_m \in {\cal C}_{2(j+e+Nm)+1}^{(q-1,q^2)}$, for $m=0,1,\cdots,\frac{t}{2}-1$. Thus, in a quite similar way as before, and by using now (\ref{pr2eq2}), (\ref{pr2eq3}), and (\ref{rm3eq3}), we have

\begin{eqnarray}
w_{\rm cplt}(\mathbf{c}_{N}(a))&=&\sum_{m=0}^{\frac{t}{2}-1}\left(w_{\rm cplt}(\mathbf{c}_{q-1}(a_m))+w_{\rm cplt}(\mathbf{c}_{q-1}(a'_m))\right)\;,  \nonumber \\
&=&\sum_{m=0}^{\frac{t}{2}-1} \sigma^{j+Nm}(W_0|W_0)+\sigma^{j+e+Nm}(W_1|W_1)\;,  \nonumber \\
&=&{\cal R}(\sigma^j({\cal G}(W_0+\sigma^e(W_1),N)))\;. \nonumber
\end{eqnarray}

\noindent
In consequence (\ref{teo3eq3}) now follows from the fact that $0\leq j<N$ and $|{\cal C}_{i}^{(N,q^2)}|=\frac{q^2-1}{N}=n$.
\end{proof}

\begin{example}\label{ejcinco}
$ $
\begin{enumerate}
\item[{\rm (A)}] Let $(q,N)=(16,5)$. In Example \ref{ejtres} {\rm (A)} we found ${\cal G}(W,N)=(4,4,4,0,4)$ and in consequence 

$$\begin{array}{lcl}
\sigma^1({\cal G}(W,N))=(4,4,4,4,0)\;, & & \sigma^2({\cal G}(W,N))=(0,4,4,4,4)\;, \\
\sigma^3({\cal G}(W,N))=(4,0,4,4,4)\;, &\mbox{ and }& \sigma^3({\cal G}(W,N))=(4,4,0,4,4)\;.
\end{array}
$$

\noindent
Thus, by Theorem \ref{teotres}, we have that $\CLI_{5}$ is a $[51,2]$ one-weight irreducible cyclic code whose nonzero weight is $48$ and whose complete weight enumerator is

\vspace{-0.5mm}
\begin{eqnarray}
\mbox{\rm CWE}_{\CLI_{5}}(Z)=1+51(&&\!\!\!\!\!\!\!\!\!z_0^4z_1^4z_2^4z_3^0z_4^4\:z_5^4z_6^4z_7^4z_8^0z_9^4\:z_{10}^4z_{11}^4z_{12}^4z_{13}^0z_{14}^4+\nonumber \\
                                  &&\!\!\!\!\!\!\!\!\!z_0^4z_1^4z_2^4z_3^4z_4^0\:z_5^4z_6^4z_7^4z_8^4z_9^0\:z_{10}^4z_{11}^4z_{12}^4z_{13}^4z_{14}^0+ \nonumber \\
                                  &&\!\!\!\!\!\!\!\!\!z_0^0z_1^4z_2^4z_3^4z_4^4\:z_5^0z_6^4z_7^4z_8^4z_9^4\:z_{10}^0z_{11}^4z_{12}^4z_{13}^4z_{14}^4+ \nonumber \\
                                  &&\!\!\!\!\!\!\!\!\!z_0^4z_1^0z_2^4z_3^4z_4^4\:z_5^4z_6^0z_7^4z_8^4z_9^4\:z_{10}^4z_{11}^0z_{12}^4z_{13}^4z_{14}^4+ \nonumber \\
                                  &&\!\!\!\!\!\!\!\!\!z_0^4z_1^4z_2^0z_3^4z_4^4\:z_5^4z_6^4z_7^0z_8^4z_9^4\:z_{10}^4z_{11}^4z_{12}^0z_{13}^4z_{14}^4\:)\;. \nonumber
\end{eqnarray}

\item[{\rm (B)}] Let $(q,N)=(9,4)$. In Example \ref{ejdos} {\rm (B)} we found $W_0=(1,2,0,2)$ and $W_1=(0,2,2,0)$. Thus, 

\vspace{-0.5mm}
\begin{eqnarray}
{\cal G}(W_0,\frac{N}{2})&=&{\cal G}((1,2,0,2),2)=(1,2)+(0,2)=(1,4)\;,\;\;\;\mbox{ and} \nonumber \\
{\cal G}(W_1,\frac{N}{2})&=&{\cal G}((0,2,2,0),2)=(0,2)+(2,0)=(2,2)\;. \nonumber
\end{eqnarray}

\noindent
Clearly, $\sigma^1({\cal G}(W_0,\frac{N}{2}))=(4,1)$ and $\sigma^1({\cal G}(W_1,\frac{N}{2}))=(2,2)$. Thus, by Theorem \ref{teotres}, we have that $\CLI_{4}$ is a $[20,2]$ two-weight irreducible cyclic code whose nonzero weights are $16$ and $20$, and whose complete weight enumerator is

\vspace{-0.5mm}
\begin{eqnarray}
\mbox{\rm CWE}_{\CLI_{4}}(Z)=1+20(&&\!\!\!\!\!\!\!\!\!\!z_0^1z_1^4z_2^1z_3^4z_4^1z_5^4z_6^1z_7^4+z_0^2z_1^2z_2^2z_3^2z_4^2z_5^2z_6^2z_7^2+\nonumber \\
                                  &&\!\!\!\!\!\!\!\!\!\!z_0^4z_1^1z_2^4z_3^1z_4^4z_5^1z_6^4z_7^1+z_0^2z_1^2z_2^2z_3^2z_4^2z_5^2z_6^2z_7^2\:)\;. \nonumber
\end{eqnarray}

\item[{\rm (C)}] Let $(q,N)=(11,5)$. Therefore $e=2$. In Example \ref{ejdos} {\rm (C)} we found $W_0=(2,1,0,0,2)$ and $W_1=(0,2,2,0,2)$. Since $\sigma^2(W_1)=(0,2,0,2,2)$,

$${\cal G}(W_0+\sigma^e(W_1),N)={\cal G}((2,1,0,0,2)+(0,2,0,2,2),5)=(2,3,0,2,4)\;,$$

and clearly

$$\begin{array}{lcl}
\sigma^1({\cal G}(W_0+\sigma^e(W_1),N))=(4,2,3,0,2)\;, & & \sigma^2({\cal G}(W_0+\sigma^e(W_1),N))=(2,4,2,3,0)\;, \\
\sigma^3({\cal G}(W_0+\sigma^e(W_1),N))=(0,2,4,2,3)\;, &\mbox{ and }& \sigma^3({\cal G}(W_0+\sigma^e(W_1),N))=(3,0,2,4,2)\;.
\end{array}
$$

\noindent
Thus, by Theorem \ref{teotres}, we have that $\CLI_{5}$ is a $[24,2]$ one-weight irreducible cyclic code whose nonzero weight is $22$ and whose complete weight enumerator is

\begin{eqnarray}
\mbox{\rm CWE}_{\CLI_{5}}(Z)=1+24(&&\!\!\!\!\!\!\!\!\!\!z_0^2z_1^3z_2^0z_3^2z_4^4\:z_5^2z_6^3z_7^0z_8^2z_9^4 +\nonumber \\
                                   &&\!\!\!\!\!\!\!\!\!\!z_0^4z_1^2z_2^3z_3^0z_4^2\:z_5^4z_6^2z_7^3z_8^0z_9^2 +\nonumber \\
                                   &&\!\!\!\!\!\!\!\!\!\!z_0^2z_1^4z_2^2z_3^3z_4^0\:z_5^2z_6^4z_7^2z_8^3z_9^0 +\nonumber \\
                                   &&\!\!\!\!\!\!\!\!\!\!z_0^0z_1^2z_2^4z_3^2z_4^3\:z_5^0z_6^2z_7^4z_8^2z_9^3 +\nonumber \\
                                   &&\!\!\!\!\!\!\!\!\!\!z_0^3z_1^0z_2^2z_3^4z_4^2\:z_5^3z_6^0z_7^2z_8^4z_9^2\:)\;. \nonumber
\end{eqnarray}
\end{enumerate}
\end{example}

\section{An application to systematic authentication codes}\label{secseis}
A systematic authentication code is a four-tuple $({\cal S},{\cal T}$,\:${\cal K}$,\:$\{E_k\:|\:k\in {\cal K}\})$, where ${\cal S}$ is the source state space associated with a probability distribution, ${\cal T}$ is the {\em tag space}, ${\cal K}$ is the {\em key space}, and $E_k : {\cal S} \to {\cal T}$ is called an {\em encoding rule}. Based on a linear error-correcting code, a systematic authentication code can be constructed as follows (for a more detailed description on the authentication codes, we refer the readers to \cite{Ding1, Ding2}):

\begin{definition}\label{defultima} 
Let $\CLI$ be an $[n,k,d]$ linear code over $\bbbf_q$ and for any $0\leq i \leq q^k-1$ let $\mathbf{c}_i=(c_{i,0},c_{i,1},\cdots,c_{i,n-1})$ be a codeword of $\CLI$. Then the {\em authentication code} based on $\CLI$ is the four-tuple,

$$({\cal S},{\cal T},{\cal K},{\cal E})=(\bbbz_{q^k},\bbbf_q,\bbbz_{n}\times \bbbf_q,\{E_k\:|\:k\in {\cal K}\})\;,$$

\noindent
where, for any $s\in {\cal S}$ and any $k=(k_1,k_2)\in {\cal K}$, the encoding rule is $E_k(s)=c_{s,k_1}+k_2$.
\end{definition}

Prior to any message interchange, a transmitter and a receiver should safely share a key $k$ for authentication purpose. A message $m=(s,t)\in {\cal S}\times {\cal T}$, that is received through a public communication channel, will be considered authentic iff $t=E_k(s)$. An {\em impersonation attack} takes place when an opponent inserts a message into the channel in the hope that it will be accepted as authentic. In a {\em substitution attack} the opponent intercepts, changes and resends a message, hoping that the receiver accepts it as authentic. In what remains, we denote the {\em maximum success probability} of the impersonation attack and the substitution attack by $P_I$ and $P_S$, respectively, and we recall that it is desirable for $P_I$ and $P_S$ to be as small as possible. In that sense, the following result is quite important:

\begin{theorem}\label{teosieste}
\cite[Theorem 1]{Ding1} Let $\CLI$ be as in Definition \ref{defultima}. Then:

$$P_I=\frac{1}{q}\;\;\;\mbox{ and }\;\;\; P_S=\!\!\!\mbox{\begin{tabular}{c} $\max$ \\ {\tiny \begin{tabular}{c} $0\neq \mathbf{c}\in\CLI$ \\ $u\in \bbbf_q$ \end{tabular}} \end{tabular}} \!\!\! \frac{\Occ(u,\mathbf{c})}{n}\geq 1-\frac{d}{n}\;,$$

\noindent
where we recall that $\Occ(u,\mathbf{c})$ denotes the number of times $u$ occurs in the codeword $\mathbf{c}$.
\end{theorem}

By considering the previous theorem, we say that an authentication code based on the linear code $\CLI$, given by Definition \ref{defultima}, is {\em optimal} if $P_S=1-\frac{d}{n}$ and {\em almost optimal} if $P_S=1-\frac{d-1}{n}$. The following result shows that the authentication codes, based on some irreducible cyclic codes studied here, are optimal or almost optimal.

\begin{theorem}\label{teoocho}
Let $\CLI_{q-1}$ be as in Theorem \ref{teodos}. Then,

$$P_S=\frac{2}{q+1}\;.$$ 

\noindent
Furthermore, the authentication code, based on the irreducible cyclic code $\CLI_{q-1}$, is optimal if $q$ is odd and is almost optimal if $q$ is even.
\end{theorem}

\begin{proof}
Owing to Theorem \ref{teosieste} and Remark \ref{remtres}, 

$$P_S=\!\!\!\mbox{\begin{tabular}{c} $\max$ \\ {\tiny \begin{tabular}{c} $0\neq \mathbf{c}\in \CLI_{q-1}$ \\ $u\in \bbbf_q$ \end{tabular}} \end{tabular}} \!\!\! \frac{\Occ(u,\mathbf{c})}{n}=\!\!\!\mbox{\begin{tabular}{c} $\max$ \\ {\tiny \begin{tabular}{c} $0\neq \mathbf{c}\in \CLI_{q-1}$ \\ $u\in \bbbf_q$ \end{tabular}} \end{tabular}} \!\!\! \frac{2}{q+1}=\frac{2}{q+1}\;.$$

\noindent
Now, in accordance with Theorem \ref{teodos}, if $q$ is odd, then $\CLI_{q-1}$ is a $[q+1,2,q-1]$ two-weight irreducible cyclic code and therefore $1-\frac{d}{n}=1-\frac{q-1}{q+1}=\frac{2}{q+1}=P_S$. On the other hand, if $q$ is even, then $\CLI_{q-1}$ is a $[q+1,2,q]$ one-weight irreducible cyclic code and therefore $1-\frac{d-1}{n}=1-\frac{q-1}{q+1}=\frac{2}{q+1}=P_S$.
\end{proof}

\section{Conclusions}\label{conclusiones}

To obtain an algebraic expression for the complete weight enumerator of a linear code over a finite field $\bbbf_q$, it is first necessary to denote the elements of $\bbbf_q$ in some fixed order, and in most of the cases this is done in an arbitrary way (see for example \cite{Bae}, \cite{Chan}, \cite{Heng}, \cite{Li-Yue}, \cite[p. 141]{MacWilliams}, \cite{Yang1,Yang2,Yang3,Zheng}, and \cite{Zhu}). Unfortunately, the price we pay for doing this is that such algebraic expressions are usually complex and quite difficult to follow. In this paper, and similarly to what is done in \cite{Blake}, we denote the elements of $\bbbf_q$ by using $u_{-1}=0$, $u_0=\alpha^0=1, u_1=\alpha^1, \cdots, u_{q-2}=\alpha^{q-2}$, where $\alpha$ is a primitive element of $\bbbf_q$ (see Definition \ref{defcero}). Thus, in addition to our vectors $W$, $W_1$, and $W_2$, it is important to emphasize that this particular way of denoting the elements of $\bbbf_q$ was the key, in Theorem \ref{teotres}, to attaining relatively simple and compact expressions for the complete weight enumerators (see specifically (\ref{teo3eq1}), (\ref{teo3eq2}), and (\ref{teo3eq3})) of the family of irreducible cyclic codes over the finite field $\bbbf_q$, of dimension two and length $n=\frac{q^2-1}{N}$, where $N|(q-1)$.

In \cite{Shi} and \cite{Vega1}, the weight distribution of any irreducible cyclic code of dimension two is determined, showing that all those irreducible cyclic codes have at most two nonzero weights. In Theorem \ref{teotres} we determined the complete weight distribution of a family of irreducible cyclic codes of dimension two. Thus, as a continuation of this work and similar to Theorem \ref{teoseis}, we believe that it would be interesting to extend Theorem \ref{teotres} in order to obtain the complete weight distribution of any irreducible cyclic code of dimension two.

In Theorem \ref{teoocho}, we determined the maximum success probability of the substitution attack, $P_S$, for the authentication codes based on the irreducible cyclic codes in Theorem \ref{teodos}. With this, we also showed that these authentication codes are optimal or almost optimal. There is evidence of the existence of optimal authentication codes based on the irreducible cyclic codes in Theorem \ref{teotres}. For example, through direct inspection of Example \ref{ejcinco} (B), it is not difficult to see that the authentication code based on the irreducible cyclic code $\CLI_{4}$ over $\bbbf_9$ is optimal with $P_S=2/10$. Thus, we also believe that it would be interesting to obtain the values of $P_S$ for the authentication codes based on the irreducible cyclic codes in Theorem \ref{teotres}.

\bibliographystyle{IEEE}
\bibliography{refs}

\end{document}